\documentclass[a4paper,11pt]{article}
\pdfoutput=1 

\usepackage{jheppub} 

\usepackage[T1]{fontenc} 

\usepackage{graphicx}
\usepackage{epstopdf}
\usepackage{amsfonts}
\usepackage{amsmath}
\usepackage{bbold}
\usepackage{booktabs}
\usepackage{siunitx}
\usepackage{caption}
\usepackage{amssymb}
\usepackage{latexsym}
\usepackage{hyperref}

\usepackage{subcaption}
\usepackage[labelformat=parens,labelsep=quad,skip=3pt]{caption}
\usepackage{graphicx}

\begin{document} 
\title{\boldmath A Very Heavy Sneutrino as Viable Thermal Dark Matter
  Candidate in $U(1)'$ Extensions of the MSSM}

\author[a]{Manuel Drees,}
\affiliation[a]{BCTP and Physics Institute, University of Bonn, Nussallee 12, D-53115 Bonn, Germany}

\author[a,b]{Felipe A. Gomes Ferreira}
\affiliation[b]{Centro Brasileiro de Pesquisas F\'{i}sicas (CBPF), Rua Dr. Xavier Sigaud 150, Urca, Rio de Janeiro, CEP 22290-180, Brazil}

\abstract{We study the Standard Model singlet (``right-handed'')
  sneutrino $\tilde \nu_R$ dark matter in a class of $U(1)'$
  extensions of the MSSM that originate from the breaking of the $E_6$
  gauge group. These models, which are referred to as UMSSM, contain
  three right--handed neutrino superfields plus an extra gauge boson
  $Z'$ and an additional SM singlet Higgs with mass $\simeq M_{Z'}$,
  together with their superpartners. In the UMSSM the right sneutrino
  is charged under the extra $U(1)'$ gauge symmetry; it can therefore
  annihilate via gauge interactions. In particular, for
  $M_{\tilde \nu_R} \simeq M_{Z'}/2$ the sneutrinos can annihilate by
  the exchange of (nearly) on--shell gauge or Higgs bosons. We focus
  on this region of parameter space. For some charge assignment we
  find viable thermal $\tilde \nu_R$ dark matter for mass up to
  $\sim 43$ TeV. This is the highest mass of a good thermal dark
  matter candidate in standard cosmology that has so far been found in
  an explicit model. Our result can also be applied to other models of
  spin$-0$ dark matter candidates annihilating through the resonant
  exchange of a scalar particle. These models cannot be tested at the
  LHC, nor in present or near--future direct detection experiments,
  but could lead to visible indirect detection signals in future
  Cherenkov telescopes.}

\emailAdd{felipeagf@cbpf.br}
\emailAdd{drees@th.physik.uni-bonn.de}

\maketitle
\flushbottom

\section{Introduction}
\label{sec:intro}

Supersymmetry (SUSY) is one of the best motivated theories to describe
new physics beyond the Standard Model (SM) at the TeV scale. It
introduces a space-time symmetry that relates bosons and fermions
which can be used to cancel the quadratic divergences that appears in
the radiative corrections of the masses of scalar bosons, providing
thus a natural solution to the hierarchy problem of the SM. It allows
the gauge couplings to unify at a certain grand unified scale in the
vicinity of the Planck scale \cite{Amaldi:1991cn, Ellis:1990wk,
  Langacker:1991an} and this can be seen as a clear hint that SUSY is
the next step towards a grand unified theory (GUT)
\cite{Mohapatra:1999vv}.

One of the most interesting features of low energy supersymmetric
models with conserved $R$ parity is that the lightest supersymmetric
particle (LSP) is absolutely stable and behaves as a realistic weakly
interacting massive particle (WIMP) dark matter (DM) candidate
\cite{Ellis:1983ew, Jungman:1995df}. Since WIMPs have non--negligible
interactions with SM particles, they can be searched for in a variety
of ways. Direct WIMP search experiments look for the recoil of a
nucleus after elastic WIMP scattering. These experiments have now
begun to probe quite deeply into the parameter space of many WIMP
models \cite{pdg, Aprile:2018dbl}. The limits from these experiments
are strongest for WIMP masses around 30 to 50 GeV. For lighter WIMPs
the recoil energy of the struck nucleus might be below the
experimental threshold, whereas the sensitivity to heavier WIMPs
suffers because their flux decreases inversely to the mass.

It is therefore interesting to ask how heavy a WIMP can be. As long as
no positive WIMP signal has been found, an upper bound on the WIMP
mass can only be obtained within a specific production mechanism,
i.e. within a specific cosmological model. In particular, nonthermal
production from the decay of an even heavier, long--lived particle can
reproduce the correct relic density for any WIMP mass, if the mass,
lifetime and decay properties of the long--lived particle are chosen
appropriately \cite{Gelmini:2006pw}. Here we stick to standard
cosmology, where the WIMP is produced thermally from the hot gas of SM
particles. The crucial observation is that the resulting relic density
is inversely proportional to the annihilation cross section of the WIMP
\cite{kt}. It has been known for nearly thirty years that the unitarity
limit on the WIMP annihilation cross section leads to an upper bound on
its mass \cite{Griest:1989wd}. Using the modern determination of the
DM density \cite{Ade:2015xua},
\begin{equation} \label{relden}
\Omega_{\rm DM} h^2 = 0.1188\pm 0.0010\,,
\end{equation}
the result of \cite{Griest:1989wd} translates into the upper bound
\begin{equation} \label{unitarity}
m_\chi \leq 120 \ {\rm TeV}\,.
\end{equation}

While any elementary WIMP $\chi$ has to obey this bound, it is not
very satisfying. Not only is the numerical value of the bound well
above the range that can be probed even by planned colliders; a
particle that interacts so strongly that the annihilation cross
section saturates the unitarity limit can hardly be said to qualify as
a WIMP. In order to put this into perspective, let us have a look at
the upper bound on the WIMP mass in specific models.

An $SU(2)$ non--singlet WIMP can annihilate into $SU(2)$ gauge bosons
with full $SU(2)$ gauge strength. For a spin$-1/2$ fermion and using
tree--level expressions for the cross section, this will reproduce the
desired relic density (\ref{relden}) for $m_\chi \simeq 1.1$ TeV for a
doublet (e.g., a higgsino--like neutralino in the MSSM
\cite{Edsjo:1997bg}); about $2.5$ TeV for a triplet (e.g., a
wino--like neutralino in the MSSM \cite{Edsjo:1997bg}); and $4.4$ TeV
for a quintuplet \cite{Cirelli:2005uq}. Including large one--loop
(``Sommerfeld'') corrections increases the desired value of the quintuplet
mass to about $9.6$ TeV \cite{Cirelli:2009uv}.

One way to increase the effective WIMP annihilation cross section is
to allow for co--annihilation with strongly interacting particles
\cite{Griest:1990kh}. Co--annihilation happens if the WIMP is close in
mass to another particle $\chi'$, and reactions of the kind
$\chi + f \leftrightarrow \chi' + f'$, where $f, f'$ are SM
particles, are not suppressed. In this case $\chi \chi'$ and
$\chi'\chi'$ annihilation reactions effectively contribute to the
$\chi$ annihilation cross section. If $\chi'$ transforms
non--trivially under $SU(3)_C$, the $\chi' \chi'$ annihilation cross
section can be much larger than that for $\chi \chi$ initial
states. On the other hand, $\chi'$ then effectively also counts as
Dark Matter, increasing the effective number of internal degrees of
freedom of $\chi$. For example, in the context of the MSSM,
co--annihilation with a stop squark \cite{Boehm:1999bj} can allow even
$SU(2)$ singlet (bino--like) DM up to about $3.3$ TeV
\cite{Harz:2018csl}, or even up to $\sim 6$ TeV if the mass splitting
is so small that the lowest stoponium bound state has a mass below
twice that of the bino \cite{Biondini:2018pwp}. Co--annihilation with
the gluino \cite{Profumo:2004wk} can put this bound up to $\sim 8$ TeV
\cite{Ellis:2015vaa}. Very recently it has been pointed out that
nonperturbative co--annihilation effects after the QCD transition
might allow neutralino masses as large as $100$ TeV if the mass splitting
is below the hadronic scale \cite{coan_new}; the exact value of the
bound depends on non--perturbative physics which is not well under
control.

The WIMP annihilation cross section can also be greatly increased if
the WIMP mass is close to half the mass of a potential $s-$channel
resonance $R$. Naively this can allow the cross section to
(nearly) saturate the unitarity limit, if one is right on resonance.
In fact the situation is not so simple \cite{Griest:1990kh}, since the
annihilation cross section has to be thermally averaged: because WIMPs
still have sizable kinetic energy around the decoupling temperature,
this average smears out the resonance. In the MSSM the potentially
relevant resonances for heavy WIMPs are the heavy neutral Higgs
bosons; in particular, neutralino annihilation through exchange of the
CP--odd Higgs $A$ can occur from an $S-$wave initial state
\cite{Drees:1992am}. However, the neutralino coupling to Higgs bosons
is suppressed by gaugino--higgsino mixing; it will thus only be close
to full strength if the higgsino and gaugino mass parameters are {\em
  both} close to $M_A/2$.

In this paper we therefore focus on models where the MSSM is extended
by an extra $U(1)'$ group, yielding the ``UMSSM''. This not only gives
rise to a new gauge boson $Z'$, but also to an additional Higgs field
$s$ whose vacuum expectation value (VEV) breaks the additional
$U(1)'$. This can provide a natural solution to the $\mu$ problem of
the MSSM \cite{Kim:1983dt} where the $\mu$ term is generated
dynamically by the VEV of $s$ \cite{Cvetic:1997ky}. Although this
solution is similar to the one provided by the next-to-minimal
supersymmetric standard model (NMSSM) \cite{Ellwanger:2009dp}, the
UMSSM is free of the cosmological domain wall problem because the
$U(1)'$ symmetry forbids the appearance of domain walls which are
created by the $Z_{3}$ discrete symmetry of the NMSSM
\cite{Han:2004yd}. 

For concreteness we work in the $E_6$ inspired version of the UMSSM.
Models of this kind were first studied more than 30 years ago in the
wake of the first ``superstring revolution'' \cite{London:1986dk,
  Hewett:1988xc}. This framework allows to study a wide range of
$U(1)'$ groups, since $E_6$ contains {\em two} $U(1)$ factors beyond
the SM gauge group.

We also add three SM singlet right--handed neutrino superfields
$\hat N^C_i$ to the spectrum. Their fermionic members are needed to cancel
anomalies related to the $U(1)'$. Moreover, the scalar members
$\tilde \nu_{R,i}$ of these superfields make good WIMP candidates
\cite{Lee:2007mt, Cerdeno:2008ep, Allahverdi:2009ae, Khalil:2011tb,
  Belanger:2011rs, Belanger:2015cra, Belanger:2017vpq}. This is in
contrast with the left--handed sneutrinos of the MSSM, which have been
ruled out as DM candidates by direct WIMP searches because their
scattering cross sections on nuclei are too large
\cite{Falk:1994es}. Right--handed sneutrinos have small scattering
cross sections on nuclei. Moreover, being scalar $SU(2)$ singlets, a
right--handed sneutrino only has two degrees of freedom; in contrast,
a higgsino--like neutralino, which also has unsuppressed couplings to
the $Z'$ boson in many cases, effectively has eight (an $SU(2)$
doublet of Dirac fermions, once co--annihilation has been included).

While the new Higgs superfield $\hat S$ is a singlet under the SM
gauge group, it is charged under $U(1)'$. This forbids an $\hat S^3$
term in the superpotential. Hence the quartic scalar interaction of
this field is determined uniquely by its $U(1)'$ charge. As a result,
the mass of the physical, CP--even Higgs boson $h_3$ is {\em
  automatically} very close to that of the $Z'$ boson, in the relevant
limit $M_{Z'} \gg M_Z$. Hence for
$M_{\tilde \nu_{R,1}} \simeq M_{Z'}/2$ the annihilation cross section
of the lightest right--handed sneutrino $\tilde \nu_{R,1}$ is enhanced
by {\em two} resonances. Out of those, the exchange of $h_3$ is more
important since it can be accessed from an $S-$wave initial state.
For a complex scalar, $Z'$ exchange is accessible only from a $P-$wave
initial state, which suppresses the thermally averaged cross section.
Notice that the $h_3 \tilde \nu_{R,i} \tilde \nu^*_{R,i}$ coupling
contains terms that are proportional to the VEV of $s$, which sets the
scale of the $Z'$ mass; for $M_{\tilde \nu_{R,1}} \simeq M_{Z'}/2$
this dimensionful coupling therefore does not lead to a suppression of
the cross section. Finally, the couplings of $h_3$ to the doublet
Higgs bosons can be tuned by varying a trilinear soft breaking
term. This gives another handle to maximize the thermally averaged
$\tilde \nu_{R,1}$ annihilation cross section in the resonance region.

The remainder of this paper is organized as follows. In
section~\ref{section2} we describe the theoretical framework of the
UMSSM and discuss its particle content, with a particular emphasis on
the gauge, Higgs, sneutrino and neutralino sectors. We describe the
calculation of the relic density, and explain our procedure to minimize it in section~\ref{section3}. In section~\ref{section4} we
first present the results of our numerical analysis for two specific
$U(1)'$ models derived from $E_6$ and then, after following the same
procedure in other $U(1)'$ models, we show the distribution of the DM
upper limit of the RH sneutrino mass in the whole UMSSM; prospects of
probing such scenarios experimentally are also discussed. Finally,
section~\ref{section5} summarizes and concludes the paper.

\section{The UMSSM}  
\label{section2}

\subsection{Model Description}  
\label{subsection2.1}

We focus on Abelian extensions of the MSSM with gauge group
$SU(3)_{C}\times SU(2)_{L}\times U(1)_{Y}\times U(1)'$, which can
result from the breaking of the $E_6$ gauge symmetry
\cite{London:1986dk, Hewett:1988xc}. In other words, it can be seen as
the low energy limit of a -- possibly string-inspired -- $E_6$ grand
unified gauge theory. $E_6$ contains $SO(10) \times U(1)_\psi$ and,
since $SO(10)$ can be decomposed into $SU(5)\times U(1)_\chi$ where $SU(5)$ 
contains the entire gauge group of the SM, one can break $E_6$ into
$SU(3)_C \times SU(2)_L \times U(1)_Y \times U(1)_\psi \times
U(1)_\chi$. Here we assume that only one extra $U(1)$ factor survives
at the relevant energy scale, which in general is a linear
combination of $U(1)_\psi$ and $U(1)_\chi$, parameterized
by a mixing angle $\theta_{E_{6}}$ \cite{London:1986dk}
\begin{equation} \label{eq 1.1}
U(1)' = \sin \theta_{E_6} U(1)'_\psi + \cos \theta_{E_6} U(1)'_\chi\,,
\end{equation}
with $\theta_{E_6} \in [-\frac{\pi}{2},\frac{\pi}{2}]$. The $U(1)'$
charges of all the fields contained in the model are then given by
\begin{equation} \label{eq 1.2}
Q'(\theta_{E_6}) = \sin \theta_{E_6} Q'_\psi + \cos \theta_{E_6} Q'_\chi\,,
\end{equation}
where $Q'_\psi$ and $Q'_\chi$ are the charges associated to the gauge
groups $U(1)'_\psi$ and $U(1)'_\chi$, respectively. In addition to the
new vector superfield $\hat{B'}$ and the MSSM superfields, the UMSSM
contains one electroweak singlet supermultiplet
$\hat{S}\equiv (s,\tilde{s})$, with a scalar field $s$ that breaks the
$U(1)'$ gauge symmetry, and three RH neutrino supermultiplets
$\hat{N}^{c}_{i}\equiv (\tilde{\nu}_{R}^{c},\nu_{R}^{c})_{i}$.

\begin{table}[h]
\begin{center}
\begin{tabular}{ |c|c|c|c|c|c|c| }  
\hline
 & {$2\sqrt{6} Q'_\psi$} & {$2\sqrt{10} Q'_\chi$} & {$2\sqrt{10} Q'_N$} & 
{$2\sqrt{15} Q'_\eta$} & {$2\sqrt{15} Q'_S$} & {$2 Q'_I$}  \\ 
    \hline
$\theta_{E_6}$  & $\frac{\pi}{2}$ & 0 & $\arctan\sqrt{15}$ & 
$-\arctan\sqrt{5/3}$ & $\arctan(\sqrt{15}/9)$ & $\arctan\sqrt{3/5}$  \\ 
\hline
    $Q'_Q$  & 1  & -1 & 1  & -2 & -1/2 & 0    \\
    $Q'_{U^C}$  & 1  & -1 & 1  & -2 & -1/2 & 0   \\
    $Q'_{D^C}$  & 1  & 3 & 2  & 1 & 4  & -1      \\ 
    $Q'_L$  & 1   & 3 & 2   & 1 & 4 & -1  \\
    $Q'_{N^C}$  & 1  & -5 & 0  & -5 & -5 & 1     \\
    $Q'_{E^C}$  & 1  & -1 & 1  & -2 & -1/2 & 0    \\
    $Q'_{H_u}$  & -2  & 2 & -2  & 4 & 1  & 0      \\ 
    $Q'_{H_d}$  & -2  & -2 & -3  & 1 & -7/2 & 1   \\
    $Q'_S$ & 4   & 0 & 5   & -5 & 5/2 & -1    \\
\hline
\end{tabular}
\caption{$U(1)'$ charges of the chiral superfields contained in the UMSSM, 
for certain values of $\theta_{E_6}$.}
\label{T1}
\end{center}
\end{table}

In Table~\ref{T1} we give the $U(1)'$ charge of all relevant matter
and Higgs fields in the UMSSM for certain values of the mixing angle
$\theta_{E_6}$. It should be noted that $U(1)_\psi$ and $U(1)_\chi$
are both anomaly--free over complete (fermionic) representations of
$E_6$. Since $U(1)_\chi$ is a subgroup of $SO(10)$, which is also
anomaly--free over complete representations of $SO(10)$, and the SM
fermions plus the right--handed neutrino complete the
${\bf 16}-$dimensional representation of $SO(10)$, $U(1)_\chi$ is
anomaly--free within the fermion content we show in the
table. However, $U(1)_\psi$ will be anomaly--free only after we
include the ``exotic'' fermions that are contained in the
${\bf 27}-$dimensional representation of $E_6$, but are not contained
in the ${\bf 16}$ of $SO(10)$. Here we assume that these exotic
superfields are too heavy to affect the calculation of the
$\tilde \nu_{R,1}$ relic density. We will see that this assumption is
not essential for our result.

\begin{figure}[h]
\centering
\includegraphics[width=0.8\textwidth]{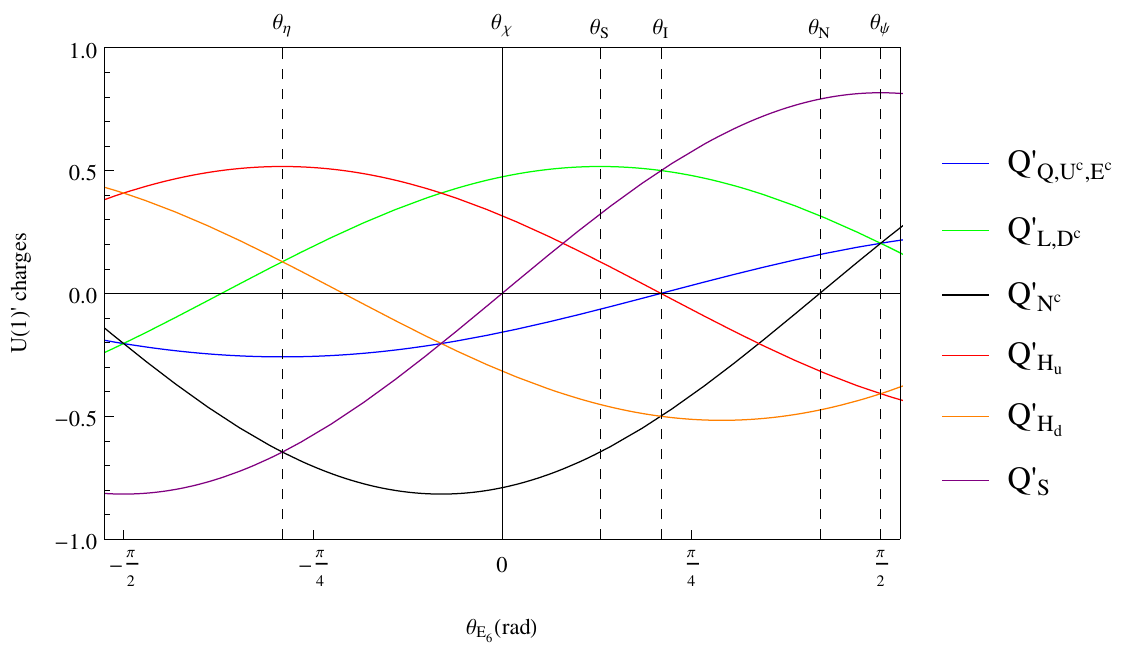}
\caption{$U(1)'$ charges of all the chiral superfields of the UMSSM as 
function of $\theta_{E_6}$.} 
\label{Fig1}
\end{figure}

Figure~\ref{Fig1} shows these charges as functions of the mixing angle
$\theta_{E_6}$. We identify by vertical lines values of $\theta_{E_6}$
that generate the well--known $U(1)'$ groups denoted by
$U(1)'_{\psi}$, $U(1)'_N$, $U(1)'_I$, $U(1)'_S$, $U(1)'_\chi$ and
$U(1)'_\eta$. The black curve in Fig.~\ref{Fig1} shows that for
$\theta_{E_6} = \arctan{\sqrt{15}}$ the $U(1)'$ charge of the RH
(s)neutrinos vanishes; this corresponds to the $U(1)'_N$ model of
Table~\ref{T1}. This model is not of interest to us, since the
$\tilde \nu_{R,i}$ are then complete gauge singlets, and do not couple
to any potential $s-$channel resonance. Similarly, for
$\theta_{E_6} = 0$, i.e.  $U(1)' = U(1)'_\chi$, the charge of $\hat S$
vanishes; in that case $s$ cannot be used to break the gauge symmetry,
i.e. the field content we have chosen is not sufficient to achieve the
complete breaking of the (extended) electroweak gauge symmetry down to
$U(1)_{\rm QED}$. All other values of $\theta_{E_6}$ are acceptable
for us.

The superpotential of the UMSSM contains, besides the MSSM
superpotential without $\mu$ term, a term that couples the extra
singlet superfield to the two doublet Higgs superfields; this term is
always allowed, since it is part of the gauge invariant ${\bf 27}^3$
of $E_6$. The superpotential also contains Yukawa couplings for
the neutrinos. We thus have:
\begin{equation} \label{eq 1.3}
\hat W = \hat{W}_{MSSM}|_{\mu=0} + \lambda \hat{S} \hat{H}_u \cdot \hat{H}_d 
+ \hat{N}^{C} {\bf Y_{\nu}} \hat{L} \cdot \hat{H}_u\,, 
\end{equation}
where $\cdot$ stands for the antisymmetric $SU(2)$ invariant product
of two doublets. The neutrino Yukawa coupling ${\bf Y_{\nu}}$ is a
$3 \times 3$ matrix in generation space and $\lambda$ is a
dimensionless coupling. Note that for $\theta_{E_6} \neq 0$ the
$U(1)'$ symmetry forbids both bilinear $\hat N^C_i \hat N^C_j$ and
trilinear $\hat S \hat N^C_i \hat N^C_j$ terms in the
superpotential. In this model the neutrinos therefore obtain pure
Dirac masses, which means that the Yukawa couplings $Y_{\nu,ij}$ must
be of order $10^{-11}$ or less; in our numerical analysis we therefore
set ${\bf Y_\nu} = 0$.

The electroweak and the $U(1)'$ gauge symmetries are spontaneously
broken when, in the minimum of the scalar potential, the real parts of
the doublet and singlet Higgs fields acquire non--zero vacuum
expectation values. These fields are expanded as
\begin{subequations} \label{vevs}
\begin{align}
\begin{split}
  H_d^0 &= \frac{1}{\sqrt{2}} \left( v_d + \phi_d + i\sigma_d \right)\,;
\end{split}\\
\begin{split}  
  H_u^0 &= \frac{1}{\sqrt{2}} \left( v_u + \phi_u + i\sigma_u \right)\,;  
\end{split}\\
\begin{split}  
  s &= \frac{1}{\sqrt{2}} \left( v_s + \phi_s + i\sigma_s \right)\,.            
\end{split}
\end{align}
\end{subequations}
We define $\tan{\beta} = \frac{v_u} {v_d}$ and
$v = \sqrt{v_d^2 + v_u^2}$ exactly as in the MSSM; this describes the
breaking of the $SU(2)_L \times U(1)_Y$ symmetry, and makes subleading
contributions to the breaking of $U(1)'$. The latter is mostly accomplished
by the VEV of $s$. The coupling $\lambda$ in
eq.(\ref{eq 1.3}) then generates an effective $\mu-$term:
\begin{equation} \label{eq 1.4}
\mu_{\rm eff} = \lambda \frac{v_{s}}{\sqrt{2}}\,. 
\end{equation}

As well known, supersymmetry needs to be broken. We parameterize this by
soft breaking terms \cite{book}:
\begin{align} \label{eq 1.66}
-\mathcal{L}_{SB} &= m^2_{H_d} |H_d|^2 + m^2_{H_u} |H_{u}|^2 + m_S^2 |s|^2
+ \tilde{Q}^\dagger {\bf m^2_{\tilde Q}} \tilde{Q} + 
\tilde{d}_R^{\dagger} {\bf m^2_{\tilde D^C}} \tilde{d}_R \nonumber \\
&+\tilde{u}_{R}^\dagger {\bf m^2_{\tilde U^C}} \tilde{u}_R
+\tilde{L}^\dagger {\bf m^2_{\tilde L}} \tilde{L} 
+\tilde{e}_R^\dagger {\bf m^2_{\tilde{E}^{C}}} \tilde{e}_R
+\tilde{\nu}_{R}^\dagger {\bf m^2_{\tilde N^C}} \tilde{\nu}_{R} \nonumber \\
&+\frac{1}{2}\Big( M_1 \lambda_{\tilde B} \lambda_{\tilde B} + 
M_2 \lambda_{\tilde W} \lambda_{\tilde W} + M_3 \lambda_{\tilde g} \lambda_{\tilde g}
+ M_4\lambda_{\tilde B'} \lambda_{\tilde B'} + h.c.\Big) \\
&+ \Big( \tilde{u}_R^C {\bf T_u} \tilde{Q}_L \cdot H_u 
- \tilde{d}_R^C {\bf T_d} \tilde{Q}_L \cdot H_d
- \tilde{e}_R^C{\bf T_e} \tilde{L}_L \cdot H_d + T_\lambda s H_u \cdot H_d
+ \tilde{\nu}_R^C {\bf T_\nu} \tilde{L}_L \cdot H_u + h.c. \Big)\,. \nonumber
\end{align}
Here we have used the notation of {\tt SPheno} \cite{Porod:2003um,
  Porod:2011nf}. The soft scalar masses and the soft trilinear
parameters of the sfermions are again  $3 \times 3$ matrices in generation
space. In the UMSSM, the $B\mu$ term of the MSSM is induced by the
$T_\lambda$ term after the breaking of the $U(1)'$ gauge symmetry.

In the following subsections we discuss those parts of the spectrum in
a bit more detail that are important for our calculation. These are
the sfermions, in particular sneutrinos; the massive gauge bosons; the
Higgs bosons; and the neutralinos. The lightest right--handed
sneutrino is assumed to be the LSP, which annihilates chiefly through
the exchange of massive gauge and Higgs bosons in the
$s-$channel. Requiring the lightest neutralino to be sufficiently
heavier than the lightest right--handed sneutrino gives important
constraints on the parameter space. The mass matrices in these
subsections have been obtained with the help of the computer code {\tt
  SARAH} \cite{Staub:2008uz, Staub:2013tta, Staub:2015kfa}; many of
these results can also be found in refs.~\cite{Belanger:2011rs,
  Belanger:2015cra, Belanger:2017vpq}.

\subsection{Sfermions}
\label{section2.2}

In the UMSSM, the $U(1)'$ gauge symmetry induces some new $D-$term
contributions to the masses of all sfermions with nonvanishing $U(1)'$
charges.  These modify the diagonal entries of the MSSM sfermion mass
matrices:
\begin{equation} \label{eq 2.18}
\Delta_{F} = \frac{1}{2} g'^2 Q'_F \Big( Q'_{H_d} v_d^2  + Q'_{H_u} v_u^2  
+ Q'_S v_s^2 \Big)\,, 
\end{equation}
where $g'$ is the $U(1)'$ gauge coupling and
$F\in \{Q,L,D^C,U^C,E^C,N^C\}$. LHC searches for $Z'$ production in
the dilepton channel imply \cite{pdg} $M^2_{Z'} \gg M^2_Z$, and hence
$v_s^2 \gg v_u^2, \, v_d^2$. The first two terms on the right--hand
side (RHS) of eq.(\ref{eq 2.18}) are therefore essentially
negligible. However, due to the contribution $\propto v_s^2$ these
$D-$terms can dominate the sfermion masses. Moreover, depending on the
value of $\theta_{E_6}$ these terms can be positive or negative. For
example, Fig.~\ref{Fig1} shows that for
$\arctan\sqrt{15} < \theta_{E_6} < \frac{\pi}{2}$ all the sfermion
masses receive positive corrections, the corrections to the RH
sneutrino masses being the smallest ones. In contrast, for
$0 < \theta_{E_6} < \arctan\sqrt{15}$ the $D-$term contribution to the
RH sneutrino masses is negative. For $\theta_{E_6} < 0$ the RH
sneutrino masses again receive positive corrections from this
$D-$term.

The tree--level sneutrino mass matrix written in the basis 
$\left(\tilde{\nu}_L, \tilde{\nu}_R\right)$ is
\begin{equation} \label{eq 2.19}
\mathcal{M}^2_{\tilde \nu} = \left( 
\begin{array}{cc}
{\bf m}_{\tilde{\nu}_L \tilde{\nu}_L^*}^2 &-\frac{1}{2} v_d v_s \lambda 
{\bf Y_\nu^*}  + \frac{1}{\sqrt{2}} v_u {\bf T_\nu^*} \\ 
-\frac{1}{2} v_d v_s \lambda {\bf Y}_\nu^T  + \frac{1}{\sqrt{2}} v_u 
{\bf T}_\nu^T  &{\bf m}_{\tilde \nu_R\tilde \nu_R^*}^2\end{array} 
\right)\,.  
\end{equation}
The $3 \times 3$ sub--matrices along the diagonal are given by: 
\begin{subequations} \label{sneumass}
\begin{align} 
{\bf m}_{\tilde{\nu}_L\tilde{\nu}_L^*}^2 &= \Big[ \Delta_L + \frac{1}{8} 
\Big( g_1^2 + g_2^2 \Big) \Big( v_d^2 - v_u^2 \Big) \Big] {\bf 1}
+ \frac{1}{2} v_u^2 {{\bf Y}_\nu^* {\bf Y}_\nu^T} + {\bf m^2_{\tilde L}} \, ;\\ 
{\bf m}_{\tilde{\nu}_R\tilde{\nu}_R^*}^2 &= \Delta_{N^C} {\bf 1} + 
\frac{1}{2} v_u^2 {{\bf Y}_\nu^T {\bf Y}_\nu^*}  + {\bf m^2_{\tilde{N}^C}}\,,
\end{align}                  
\end{subequations}
where $g_1$ and $g_2$ are the $U(1)_Y$ and $SU(2)_L$ gauge couplings,
respectively. As noted earlier, the neutrino Yukawa couplings have to
be very small. We therefore set ${\bf Y_\nu} = {\bf T_\nu} = 0$, so
that the $6 \times 6$ matrix (\ref{eq 2.19}) decomposes into two
$3 \times 3$ matrices.\footnote{Strictly speaking some neutrino Yukawa
  couplings have to be nonzero in order to generate the required
  sub--eV neutrino masses. However, the $\tilde \nu_L - \tilde \nu_R$
  mixing induced by these tiny couplings is completely negligible for
  our purposes.} Since all interactions of the $\tilde \nu_R$ fields
are due to $U(1)'$ gauge interactions which are the same for all
generations, we can without loss of generality assume that the matrix
${\bf m^2_{\tilde{N}^C}}$ of soft breaking masses is diagonal. The
physical masses of the RH sneutrinos are then simply given by
$m^2_{\tilde \nu_{R,i}} = m^2_{\tilde N^C_i} + \Delta_{N^C}$. Our LSP
candidate is the lightest of the three $\tilde \nu_R$ states, which we
call $\tilde \nu_{R,1}$.

\subsection{Gauge Bosons}
\label{section2.3}

The UMSSM contains three neutral gauge bosons, from the $SU(2)_L, \, U(1)_Y$
and $U(1)'$, respectively. As in the SM and MSSM, after symmetry breaking
one linear combination of the neutral $SU(2)_L$ and $U(1)_Y$ gauge
bosons remains massless; this is the photon. The orthogonal state $Z_0$ mixes
with the $U(1)'$ gauge boson $Z_0'$ via a $2 \times 2$ mass matrix:
\begin{equation} \label{MM}
\mathcal{M}^2_{ZZ'}= \, \left( 
\begin{array}{cc} 
M_{Z_0}^2 & \Delta_Z \\  \Delta_Z  & M_{Z_0^{\prime}}^2 \end{array} 
\right)\,,
\end{equation}
with
\begin{eqnarray}
M_{Z_0}^2 & = & \frac{1}{4} (g_1^2+g_2^2) v^2\,;       \label{37}
\\
\Delta_Z & = & \frac{1}{2} g' \sqrt{g_1^2+g_2^2} \left( 
Q'_{H_d} v_d^2 - Q'_{H_u} v_u^2\right)\,;  \label{42}
\\
M_{Z'_0}^2 & = & g'^2 \Big( Q'^2_{H_d} v_d^2  + Q'^2_{H_u} v_u^{2}  
+ Q'^2_S v_s^2 \Big)\,.           \label{eq 2.10}
\end{eqnarray}
Recall that $g_2, \, g_1$ and $g'$ are the gauge couplings associated
to $SU(2)_L, \, U(1)_Y$ and $U(1)'$, respectively. The eigenstates
$Z$ and $Z'$ of this mass matrix can be written as:
\begin{eqnarray}
Z & = & \cos{\alpha_{ZZ'}} Z_0 + \sin{\alpha_{ZZ'}} Z_0'\,; \nonumber \label{44}
\\
Z' & = & -\sin{\alpha_{ZZ'}} Z_0 + \cos{\alpha_{ZZ'}}Z_0'\,.     \label{45}
\end{eqnarray}
The mixing angle $\alpha_{ZZ'}$ is given by
\begin{equation}
\sin{2\alpha_{ZZ'}} = \frac{2\Delta_{Z}} { M_Z^2 - M_{Z'}^2 }\,.
\end{equation}
The masses of the physical states are
\begin{equation}
M_{Z,Z'}^2 = \frac{1}{2} \left[ M_{Z_0}^2 + M_{Z'_0}^2 \mp 
\sqrt{\left( M_{Z'_0}^2 - M_{Z_0}^2 \right)^2 + 4\Delta_Z^2} \ \right]\,.
\end{equation}

Note that the off--diagonal entry $\Delta_Z$ in eq.(\ref{MM}) is of
order $v^2$. We are interested in $Z'$ masses in excess of $10$ TeV,
which implies $v_s^2 \gg v^2$. The mixing angle $\alpha_{ZZ'}$ is
${\cal O}(M_Z^2 / M^2_{Z'})$, which is automatically below current
limits \cite{pdg} if $M_{Z'} \geq 10$ TeV. Moreover, mass mixing
increases the mass of the physical $Z'$ boson only by a term of order
$M^4_Z/M^3_{Z'}$, which is less than $0.1$ MeV for $M_{Z'} \geq 10$
TeV. To excellent approximation we can therefore identify the physical
$Z'$ mass with $M_{Z'_0}$ given in eq.(\ref{eq 2.10}), with the last
term $\propto v_s^2$ being the by far dominant one.

Recall from the discussion of the previous subsection that the mass of
the right--handed sneutrinos can get a large positive contribution
from the $U(1)'$ $D-$term for some range of $\theta_{E_6}$. In fact,
from eqs.(\ref{eq 2.18}) and (\ref{eq 2.10}) together with the charges
listed in Table~\ref{T1} we find that this $D-$term contribution
exceeds $(M_{Z'}/2)^2$ if
\begin{equation} \label{th_range}
-\sqrt{15} < \tan \theta_{E_6} < 0\,.
\end{equation}
For this range of $\theta_{E_6}$ one therefore needs a negative
squared soft breaking contribution $m^2_{\tilde N^C_1}$ in order to
obtain $M_{\tilde \nu_{R,1}} \simeq M_{Z'}/2$.

Note that we neglect kinetic $Z - Z'$ mixing \cite{Kalinowski:2008iq,
  Belanger:2017vpq}. In the present context this is a loop effect
caused by the mass splitting of members of the ${\bf 27}$ of
$E_6$. This induces small changes of the couplings of the physical
$Z'$ boson, which have little effect on our result; besides, this loop
effect should be treated on the same footing as other one--loop
corrections.

\subsection{The Higgs Sector}
\label{section2.4}

The Higgs sector of the UMSSM contains two complex $SU(2)_{L}$ doublets
$H_{u,d}$ and the complex singlet $s$. Four degrees of freedom get
``eaten'' by the longitudinal components of $W^\pm, \, Z$ and
$Z'$. This leaves three neutral CP--even Higgs bosons
$h_{i},\, i\in \{1,2,3 \}$, one CP-odd Higgs boson $A$ and two charged
Higgs bosons $H^{\pm}$ as physical states. After solving the
minimization conditions of the scalar potential for the soft breaking
masses of the Higgs fields, the symmetric $3 \times 3$ mass matrix for
the neutral CP--even states in the basis $(\phi_d,\phi_u,\phi_s)$ has
the following tree--level elements:
\begin{subequations} \label{eq:subeqns}
\begin{align}
\left( \mathcal{M}_+^0 \right)_{\phi_d \phi_d} &= \Big[ \frac{g_1^2 + g_2^2} {4}
+ (Q'_{H_d})^2 g'^2 \Big] v_d^2 + \frac{T_\lambda v_s v_u} {\sqrt{2} v_d} \,;
\label{eq 2.14a} \\  
\left( \mathcal{M}_+^0 \right)_{\phi_d \phi_u} &= -\Big[ \frac{g_1^2 +g_2^2} {4}
- g'^2 Q'_{H_d} Q'_{H_u} - \lambda^2\Big] v_dv_u - \frac{T_\lambda v_s} 
{\sqrt{2}}\,;
\label{eq 2.14b} \\
\left( \mathcal{M}_+^0 \right)_{\phi_d \phi_s} &= \Big[ g'^2 Q'_{H_d} Q'_S
+ \lambda^2\Big] v_d v_s - \frac{T_\lambda v_u} {\sqrt{2}}\,;
\label{eq 2.14c} \\
\left( \mathcal{M}_+^0 \right)_{\phi_u \phi_u} &= \Big[ \frac{g_1^2 +g_2^2} {4}
+ (Q'_{H_u})^2g'^2\Big] v_u^2 + \frac{T_\lambda v_s v_d} {\sqrt{2}v_u}\,;
\label{eq 2.14d}   \\
\left( \mathcal{M}_+^0 \right)_{\phi_u \phi_s} &= \Big[ g'^2 Q'_{H_u} Q'_S
+ \lambda^2 \Big] v_u v_s - \frac{ T_\lambda v_d} {\sqrt{2}}\,;
\label{eq 2.14e}   \\
\left( \mathcal{M}_+^0 \right)_{\phi_s \phi_s} &= g'^2 (Q'_S)^2 v_s^2 
+ \frac {T_\lambda v_d v_u }{ \sqrt{2} v_s }\,.  \label{eq 2.14f}
\end{align}
\end{subequations}
In general the eigenstates and eigenvalues of this mass matrix have to
be obtained numerically. We denote the mass eigenstates by
$h_{1},\,h_{2},\,h_{3}$, ordered in mass.

The tree level mass of the single physical neutral CP--odd state is
\begin{equation} \label{eq 2.16}
M_A^2|_{\text{tree}} = \frac {\sqrt{2} T_\lambda} {\sin{2\beta}} v_s 
\left( 1 + \frac {v^2} {4v_s^2} \sin^2{2\beta} \right)\,. 
\end{equation}
In our sign convention, $\tan\beta$ and $v_s$ are positive in the
minimum of the potential; eq.(\ref{eq 2.16}) then implies that
$T_\lambda$ must also be positive. As in the MSSM $M_A^2$ differs from
the squared mass of the physical charged Higgs boson only by terms of
order $v^2$:
\begin{equation} \label{eq 2.17}
M_{H^+}^2|_{\text{tree}} = M_{W^+}^2 + \frac{\sqrt{2} T_\lambda} {\sin{2\beta}} v_s
- \frac {\lambda^2} {2} v^2\,. 
\end{equation}
Both $A$ and $H^\pm$ are constructed from the components of $H_u$ and $H_d$,
without any admixture of $s$.

Recall that we are interested in the limit $v_s \gg v$.
Eqs.(\ref{eq:subeqns}) show that the entries mixing the $SU(2)_L$
singlet with the doublets are of order $v_s v$ or $T_\lambda v$,
whereas the diagonal mass of the singlet is of order $v_s^2$. The
mixing between singlet and doublet states is therefore
small. Moreover, we will work in the MSSM--like decoupling limit
$M_A^2 \gg M_Z^2$, which ensures that the lightest neutral CP--even
Higgs boson has couplings close to those of the SM Higgs. Its mass can
then approximately be written as \cite{King:2005jy, Belanger:2017vpq}:
\begin{eqnarray}  \label{eq 1.65}
M_{h_1}^2|_{\text{tree}} &\simeq& \frac{1}{4} ( g_1^2 + g_2^2 ) v^2 \cos^22\beta
+ \frac{1}{2} \lambda^2 v^2 \sin^22\beta + g'^2 v^2 
\left(Q'_{H_d} \cos^2\beta + Q'_{H_u} \sin^2\beta \right)^2 \nonumber \\
& - & \frac{v^2} { g'^2 (Q'_S)^2} \Big[ \lambda^2 - \frac {T_\lambda \sin^22\beta}
{\sqrt 2 v_s} + g'^2 Q'_S \left( Q'_{H_d} \cos^2\beta + Q'_{H_u} \sin^2\beta
\right) \Big]^2\,.
\end{eqnarray}
The first term on the RHS is as in the MSSM. The second term is an
$F-$term contribution that also appears in the NMSSM, while the third
term is due to the $U(1)'$ $D-$ term. These terms are positive. The
second line is due to mixing between singlet and doublet states; note
that this mixing always reduces the mass of the lighter eigenstate,
but increases the mass of the heaviest state $h_3$. As well known, the
mass of $h_1$ also receives sizable loop corrections, in particular
from the top--stop sector \cite{Okada:1990vk, Haber:1990aw}; we will
briefly discuss them below when we describe our numerical procedures.

As noted above, in the limit $v_s \gg v$ the mixing between singlet and
doublet states can to first approximation be neglected. Here we chose
the heaviest state to be (mostly) singlet. From the last eq.(\ref{eq:subeqns})
and eq.(\ref{eq 2.10}) we derive the important result
\begin{equation} \label{ms}
M^2_{Z'}|_{\text{tree}} \simeq M_{h_3}^2|_{\text{tree}} + {\cal O}(v^2)\,.
\end{equation}
Here we have assumed $|T_\lambda| \leq v_s$ because for larger values
of $|T_\lambda|$ the mass of the heavy doublet Higgs can exceed the
mass of the singlet state. As we will see, in the region of parameter
space that minimizes the $\tilde \nu_{R,1}$ relic density we need
$M_A < M_{\tilde \nu_{R,1}}$.

Eq.(\ref{ms}) leads to an $h_3 - Z'$ mass splitting of order
$M^2_Z/M_{Z'}$, which is below $1$ GeV for $M_{Z'} > 10$ TeV. Loop
corrections induce significantly larger mass splittings, with
$M_{Z'} > M_{h_3}$; however, the splitting still amounts to less than
$1\%$ in the relevant region of parameter space, which is well below
the typical kinetic energy of WIMPs in the epoch around their
decoupling from the thermal bath. We thus arrive at the important
result that $M_{\tilde \nu_{R,1}} \simeq M_{Z'}$ {\em automatically}
implies $M_{\tilde \nu_{R,1}} \simeq M_{h_3}$ in our set--up, so that
$\tilde \nu_{R,1}$ annihilation is enhanced by {\em two} nearby
resonances.

\subsection{Neutralinos}
\label{section2.5}

The neutralino sector is formed by the fermionic components of the
neutral vector and Higgs
supermultiplets. So, in addition to the neutralino sector of the MSSM,
the UMSSM has another gaugino state associated with the $U(1)'$
gauge symmetry and a singlino state that comes from the extra
scalar supermultiplet $\hat{S}$. The neutralino mass matrix written in
the basis
$\left(\lambda_{\tilde{B}}, \tilde{W}^0, \tilde{H}_d^0, \tilde{H}_u^0,
  \tilde{S}, \lambda_{\tilde{B}'}\right)$ is:
\begin{equation}  \label{eq 2.21}
\mathcal{M}_{\tilde{\chi}^0} = \left( 
\begin{array}{cccccc}
M_1 &0 &-\frac{1}{2} g_1 v_d  &\frac{1}{2} g_1 v_u  &0 &0\\ 
0 &M_2 &\frac{1}{2} g_2 v_d  &-\frac{1}{2} g_2 v_u  &0 &0\\ 
-\frac{1}{2} g_1 v_d  &\frac{1}{2} g_2 v_d  &0 &- \mu_{\rm eff}  &
- \frac{1}{\sqrt{2}} v_u \lambda  &g' Q'_{H_d} v_d \\ 
\frac{1}{2} g_1 v_u  &-\frac{1}{2} g_2 v_u  &- \mu_{\rm eff}  &0 &
- \frac{1}{\sqrt{2}} v_d \lambda  &g' Q'_{H_u} v_u \\ 
0 &0 &- \frac{1}{\sqrt{2}} v_u \lambda  &- \frac{1}{\sqrt{2}} v_d \lambda  
&0 &g' Q'_{S} v_s \\ 
0 &0 &g' Q'_{H_d} v_d  &g' Q'_{H_u} v_u  &g' Q'_{S} v_s  &M_4\end{array} 
\right)\,.   
\end{equation} 
This matrix is diagonalized by a unitary 6 $\times$ 6 matrix $N$ which
gives the mass eigenstates (in order of increasing mass)
$\tilde{\chi}^0_1, \, \tilde{\chi}^0_2, \, \tilde{\chi}^0_3,
\tilde{\chi}^0_4,\, \tilde{\chi}^0_5,\, \tilde{\chi}^0_6$
as a linear combinations of the current eigenstates. We have ignored a
possible (gauge invariant) mixed $\tilde B \tilde B'$ mass term
\cite{Suematsu:1997qt, Suematsu:1997au}.

Note that the singlet higgsino (singlino for short) $\tilde S$ and the
$U(1)'$ gaugino $\tilde B'$ mix strongly, through an entry of order
$v_s$. On the other hand, these two new states mix with the MSSM only
through entries of order $v$. Therefore the eigenvalues of the
lower--right $2 \times 2$ submatrix in eq.(\ref{eq 2.21}) are to good
approximation also eigenvalues of the entire neutralino mass
matrix. Note that the smaller of these two eigenvalues decreases with
increasing $M_4$. Requiring this eigenvalue to be larger than
$M_{\tilde \nu_{R,1}} \simeq M_{Z'}/2$ therefore implies
\begin{equation} \label{neut1}
| M_4 | < \frac{3}{2} M_{Z'}\,.
\end{equation}
Moreover, the smallest mass of the MSSM--like states should also be larger
than $M_{Z'}/2$, which implies
\begin{equation} \label{neut2}
|M_1| > \frac{1}{2} M_{Z'}; \ |M_2| > \frac{1}{2} M_{Z'}; \
|\lambda| > \frac{1}{\sqrt{2}} |Q_S g'| \,.
\end{equation}
We have used eqs.(\ref{eq 1.4}) and (\ref{eq 2.10}) in the derivation of
the last inequality.

We finally note that the chargino sector of the UMSSM is identical to
that of the MSSM, with $\mu \rightarrow \mu_{\rm eff}$.

\section{Minimizing the Relic Abundance of the 
Right-Handed Sneutrino}
\label{section3}

As described in the Introduction, we want to find the upper bound on
the mass of the lightest RH sneutrino $\tilde \nu_{R,1}$ from the
requirement that it makes a good thermal WIMP in standard cosmology.
As well known \cite{kt}, under the stated assumptions the WIMP relic
density is essentially inversely proportional to the thermal average
of its annihilation cross section into lighter particles; these can be
SM particles or Higgs bosons of the extended sector. The upper bound
on $M_{\tilde \nu_{R,1}}$ will therefore be saturated for combinations
of parameters that maximize the thermally averaged
$\tilde \nu_{R,1} \tilde \nu_{R,1}^*$ annihilation cross section. 

All relevant couplings of the RH sneutrinos are proportional to
the $U(1)'$ gauge coupling $g'$. In particular, two RH sneutrinos can
annihilate into two neutrinos through exchange of a $U(1)'$ gaugino.
This, and similar reactions where one or both particles in the initial
and final state are replaced by antiparticles, are typical electroweak
$2 \rightarrow 2$ reactions without enhancement factors. They will
therefore not allow RH sneutrino masses in the multi--TeV range.

\begin{figure}[h!]
\centering
\includegraphics[width=0.4\textwidth]{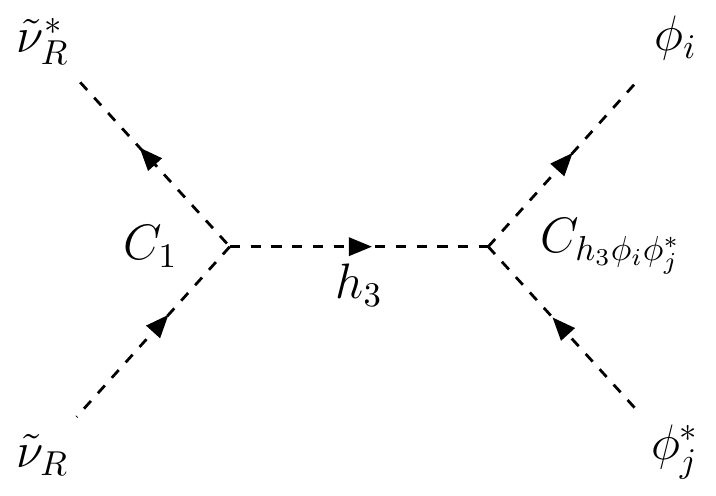}
\caption{Main annihilation process for the annihilation of RH
  sneutrinos. The final state can contain both physical Higgs
  particles and the longitudinal components of the weak $W$ and $Z$
  gauge bosons, which are equivalent to the corresponding would--be
  Goldstone modes.} \label{Fig2}
\end{figure}

In contrast, $\tilde \nu_{R,1} \tilde \nu_{R,1}^*$ annihilation
through $Z'$ and scalar $h_3$ exchange can be resonantly enhanced if
$M_{\tilde \nu_{R,1}} \simeq M_{Z'}/2$; recall that
$M_{h_3} \simeq M_{Z'}$ is automatic in our set--up, if $h_3$ is
mostly an SM singlet, as we assume. Note that the $Z'$ exchange can
only contribute if the sneutrinos are in a $P-$wave. This suppresses
the thermal average of the cross section by a factor $\geq 7$. For
comparable couplings, $h_3$ exchange, which is depicted in
Fig.~\ref{Fig2}, is therefore more important.

In the $h_3$ resonance region the annihilation cross section scales like
\begin{equation} \label{eq 3.1}
\sigma_{\rm ann} \propto \frac {(Q'_{N^C})^2} {M^2_{\tilde{\nu}_{R,1}}}\, .  
\end{equation}
Since the $h_3 \tilde \nu_{R,1} \tilde \nu_{R,1}^*$ coupling, denoted
by $C_1$ in Fig.~\ref{Fig2}, originates from the $U(1)'$ $D-$term, it
is proportional to the product of $S$ and $N^C$ charges:
\begin{equation}  \label{eq 3.3}
\left| C_1 \right| \simeq  g'^2 \left| Q'_{N^C} Q'_S v_s \right| \,.
\end{equation}
These charges are determined uniquely once the angle $\theta_{E_6}$
has been fixed. The denominator in eq.(\ref{eq 3.1}) results from
dimensional arguments, using the fact that there is essentially only
one relevant mass scale once the resonance condition has been
imposed.\footnote{The couplings $C_1$ and $C_{h_3 \phi_i \phi_j^*}$ in
  Fig.~\ref{Fig2} carry dimension of mass. They are dominated by the
  VEV $v_s$, which is proportional to $M_{Z'} \simeq M_{h_3}$, and
  hence to $M_{\tilde \nu_{R,1}}$ if the resonance condition is
  satisfied.}

Note that near the resonance the annihilation cross section is
effectively only ${\cal O}(\alpha')$, not ${\cal O}(\alpha'^2)$, where
$\alpha' = g'^2/(4\pi)$. The $\tilde \nu_{R,1} \tilde \nu_{R,1}^*$
annihilation cross section is then larger than typical
co--annihilation cross sections, if the latter are not resonantly
enhanced. Even co--annihilation with a superparticle that can also
annihilate resonantly (e.g., a higgsino--like neutralino) will not
increase the effective annihilation cross section, but will increase
the effective number of degrees of freedom per dark matter particle
$g_\chi$. As a result, we find that co--annihilation {\em reduces} the
upper bound on $M_{\tilde \nu_{R,1}}$. For example, if all three RH
sneutrinos have the same mass, the upper bound on this mass decreases
by a factor of $\sqrt{3}$, since the annihilation cross section has to
be increased by a factor of $3$ in order to compensate the increase of
$g_\chi$. We therefore require that the lightest neutralino is at
least $20\%$ heavier than $\tilde \nu_{R,1}$.

As noted earlier, the initial--state coupling $C_1$ in Fig.~\ref{Fig2}
is essentially fixed by $\theta_{E_6}$. The upper bound on
$M_{\tilde \nu_{R,1}}$ for given $\theta_{E_6}$ can therefore be found
by optimizing the final state couplings. We find that the relic
density is minimized if the effective final state coupling $C_2$,
defined more precisely below, is of the same order as $C_1$. This can
be understood as follows. For much larger values of $C_2$ the width of
$h_3$ increases, which reduces the cross section. On the other hand,
since the peak of the thermally averaged cross section is reached for
$M_{\tilde \nu_{R,1}}$ slightly below $M_{h_3}/2$
\cite{Griest:1990kh},
$h_3 \rightarrow \tilde \nu_{R,1} \tilde \nu_{R,1}^*$ decays are
allowed, and dominate the total $h_3$ width if $C_2 \ll C_1$; in this
case increasing $C_2$ will clearly increase the cross section,
i.e. reduce the relic density.

The only sizable couplings of the singlet--like Higgs state $h_3$ to
particles with even $R-$parity (i.e., to particles possibly lighter
than the LSP $\tilde \nu_{R,1}$) are to members of the Higgs doublets.
$h_3$ couples to $H_u$ and $H_d$ through the $U(1)'$ $D-$term, with
contributions $\propto g'^2 Q'_S Q'_{H_u, H_d} v_s$; through $F-$terms
associated to the coupling $\lambda$, with contributions
$\propto \lambda^2 v_s$; and through a trilinear soft breaking term,
with contributions $\propto T_\lambda$. In the decoupling limit
$M_A^2 \gg M_Z^2$ the relevant couplings are given by:
\begin{eqnarray}
C_{h_3 H^+ H^-} \simeq C_{h_3 h_2 h_2 } \simeq C_{h_3 A A} \simeq - i &\Big[& 
g'^2 \left( \cos^2\beta Q'_{H_u} + \sin^2\beta Q'_{H_d} \right) Q'_Sv_s   
\nonumber \\
&+& v_s \lambda^2 + \frac{\sin(2\beta)} {\sqrt{2}} T_\lambda \Big]\,;
 \label{eq 3.4}    \\
C_{h_3 G^+ G^-} \simeq C_{h_3 h_1 h_1} \simeq C_{h_3 G^0 G^0 } \simeq -i  &\Big[& 
g'^2 \left( \sin^2\beta Q'_{H_u} + \cos^2\beta Q'_{H_d} \right) Q'_S v_s  
\nonumber \\
& + & v_s \lambda^2 - \frac{\sin(2\beta)}{\sqrt{2}} T_\lambda \Big]\,;
 \label{eq 3.5} \\
C_{h_3 H^+ G^-} \simeq C_{h_3 h_2 h_1} \simeq C_{h_3 A G^0} \simeq -i &\Big[& 
g'^2 \frac{ \sin(2\beta)} {2} \left( Q'_{H_u} - Q'_{H_d} \right) Q'_S v_s 
\nonumber \\
& - &\frac{\cos(2\beta)}{\sqrt{2}} T_\lambda \Big]\,. \label{eq 3.6}
\end{eqnarray}
Since $M_{h_3} \gg v$, at scale $M_{h_3}$ $SU(2)_{L}$ is effectively
unbroken. The couplings of $h_3$ to two members of the heavy doublet
containing the physical states $H^\pm, h_2$ and $A$ therefore are
all the same, see eq.(\ref{eq 3.4}), as are the couplings to the light
doublet containing $h_1$ and the would--be Goldstone modes $G^0$ and
$G^\pm$, see eq.(\ref{eq 3.5}); finally, eq.(\ref{eq 3.6}) describes
the common coupling to one member of the heavy doublet and one member
of the light doublet. Of course, the would--be Goldstone modes are not
physical particles; however, again since $M_{h_3} \gg v$ the
production of physical longitudinal gauge bosons can to very good
approximation be described as production of the corresponding
Goldstone states. This is the celebrated equivalence theorem
\cite{Cornwall:1974km}.\footnote{Due to the effective restoration
of $SU(2)_{L}$ at scale $M_{h_3}$ the total decay width of $h_3$, which
determines the total annihilation cross section via $h_3$ exchange,
can still be computed from eqs.(\ref{eq 3.4}) to (\ref{eq 3.6}) even
if the decoupling limit is not reached; the dependence on the
mixing between the CP--even states drops out after summing over
all final states.}

We find numerically that the $\tilde \nu_{R,1}$ relic density is minimized
when $h_3$ decays into two members of the heavy Higgs doublet are
allowed. From eqs.(\ref{eq 2.16}) and (\ref{eq 2.17}) we see that
this requires
\begin{equation}  \label{eq 3.2}
\frac{ \sqrt{2}  T_\lambda v_s } {\sin{2\beta}} <  
\frac{1}{4} g'^2 \left( Q'_S \right)^2 v_s^2  \quad\quad\quad \Rightarrow 
\quad\quad\quad T_\lambda < \frac {g'^2 \left( Q'_S \right)^2 
\sin{2\beta}} {4 \sqrt{2}} v_s\,.
\end{equation}
This implies that the singlet--like state is indeed the heaviest physical
Higgs boson. 

We can now define an effective final--state coupling $C_2$ for the diagram
shown in Fig.~\ref{Fig2}:
\begin{equation} \label{eq 3.8}
C_2 = \sqrt{ 2 \left| C_{h_3 h_2 h_2} \right|^2 \sqrt{ 1 - 
\frac{4 M_{h_2}^2} {M_{h_3}^2} } + 2 \left| C_{h_3 h_1 h_1} \right|^2
+ 4 \left| C_{h_3 h_2 h_1} \right|^2 \left( 1 - \frac{ M_{h_2}^2} {M_{h_3}^2} 
\right)} \,.
\end{equation}
Here we have included the kinematic factors into the effective
coupling, using the same mass $M_{h_2}$ for all members of the heavy
Higgs doublet and ignoring $M_{h_1}, M_W$ and $M_Z$, which are much
smaller than $M_{h_3}$. The numerical coefficients originate from
summing over final states: $H^+ H^-, \, A A$ and $h_2 h_2$ for the
first term, where the last two final states get a factor $1/2$ for
identical final state particles; $G^+ G^-, \, G^0 G^0$ and $h_1 h_1$
for the second term, again with factor $1/2$ in front of the second
and third contribution; and $G^+ H^-,\, G^- H^+,\, G^0 A$ and
$h_1 h_2$ for the third term.

Since the contribution from $h_3$ exchange is accessible from an
$S-$wave initial state, it peaks for DM mass very close to $M_{h_3}/2$
where one needs quite small velocity to get exactly to the pole
$s=M_{h_3}^{2}$; at such a small velocity, the $Z'$ exchange
contribution, which can only be accessed from a $P-$wave initial
state, is quite suppressed. As a consequence, near the peak of the
thermally averaged total cross section the $h_3$ exchange processes
always contributes more than $90\%$ to the total, whereas the $Z'$
exchange contribution shrinks as we approach the peak. The latter
reaches its maximum at a larger difference between $M_{Z'}$ and
$2M_{\tilde \nu_{R,1}}$, but its contribution exceeds $10\%$ of the
total only if $2M_{\tilde \nu_{R,1}}$ is at least $3\%$ below $M_{Z'}$,
or else above the resonance. Note also that the annihilation into
pairs of SM fermions via $Z'$ exchange is completely determined by
$\theta_{E_6}$. In principle we could contemplate annihilation into
exotic fermions, members of $\mathbf{27}$ of $E_6$ that are
required for anomaly cancellation, as noted in Sec.~2.1. However, the
contribution from the SM fermions already sums to an effective final
state coupling which is considerably larger than the initial state
coupling; this helps to explain why the $Z'$ contribution is always
subdominant. Adding additional final states therefore reduces the $Z'$
exchange contribution to the $\tilde \nu_{R,1}$ annihilation cross
section even further. This justifies our assumption that the exotic
fermions are too heavy to affect the calculation of the
$\tilde \nu_{R,1}$ relic density.

Finally, all other processes of the model contribute at most $1\%$ to
the thermally averaged total cross section in the resonance region.
This shows that the parameters that describe the rest of the spectrum
are irrelevant to our calculation, as long as $\tilde \nu_{R,1}$ is
the LSP and sufficiently separated in mass from the other
superparticles to avoid co--annihilation. These parameters were
therefore kept fixed in the numerical results presented below.

\section{Numerical Results}  
\label{section4}

We are now ready to present numerical results. We will first describe
our procedure. Then we discuss two choices for $\theta_{E_6}$, i.e.
for the $U(1)'$ charges, before generalizing to the entire range of
possible values of this mixing angle.

\subsection{Procedure}
\label{subsection4.1}

We have used the Mathematica package {\tt SARAH} \cite{Staub:2008uz,
  Staub:2013tta, Staub:2015kfa} to generate routines for the precise
numerical calculation of the spectrum with {\tt SPheno}
\cite{Porod:2003um, Porod:2011nf}. This code calculates by default the
pole masses of all supersymmetric particles and their corresponding
mixing matrices at the full one--loop level in the
$\overline{{\rm DR}}$ scheme. {\tt SPheno} also includes in its
calculation all important two--loop corrections to the masses of
neutral Higgs bosons \cite{Goodsell:2014bna, Goodsell:2015ira,
  Braathen:2017izn}.  The dark matter relic density and the dark
matter nucleon scattering cross section relevant for direct detection
experiments are computed with {\tt micrOMEGAS-4.2.5}
\cite{Belanger:2014vza}. The mass spectrum generated by {\tt SPheno}
is passed to {\tt micrOMEGAS-4.2.5} through the SLHA+ functionality
\cite{Belanger:2010st} of {\tt CalcHep} \cite{Pukhov:2004ca,
  Belyaev:2012qa}. The numerical scans were performed by combining the
different codes using the Mathematica tool {\tt SSP}
\cite{Staub:2011dp} for which {\tt SARAH} already writes an input
template.

{\tt SARAH} can generate two different types of templates that can be
used as input files for {\tt SPheno}. One is the high scale input,
where the gauge couplings and the soft SUSY breaking parameters are
unified at a certain GUT scale and their renormalization group (RG)
evolution between the electroweak, SUSY breaking and GUT scale is
included. The other one is the low scale input where the gauge
couplings, VEVs, superpotential and soft SUSY breaking parameters of
the model are all free input parameters that are given at a specific
renormalization scale near the sparticle masses, in which case no RG
running to the GUT scale is needed. In this template the SM gauge
couplings are given at the electroweak scale and evolve to the SUSY
scale through their RGEs. The dark matter phenomenology of a model in
the WIMP context is usually well studied at low energies; moreover,
acceptable low energy phenomenology for both the $U(1)_{\psi}$ and the
$U(1)_{\eta}$ model in the limit where the singlet Higgs decouples
works much better with nonuniversal boundary conditions
\cite{Langacker:1998tc}. Finally, a bound that is valid for general
low--scale values of the relevant parameters will also hold (but can
perhaps not be saturated) in constrained scenarios.

In our work we therefore define the relevant free parameters of the
UMSSM directly at the SUSY mass scale, which is defined as the
geometric mean of the two stop masses. We created new model files for
different versions of the UMSSM to be used in {\tt SARAH} and {\tt
  SPheno} where all the $U(1)'$ charges are written in terms of the
$U(1)$ mixing angle $\theta_{E_6}$ using eq.(\ref{eq 1.1}).

Our goal is to find the upper bound on the mass of the lightest RH
sneutrino, and therefore on $M_{Z'} \simeq M_{h_3}$. We argued in
Sec.~\ref{section3} that co--annihilation would weaken the bound. We
therefore have to make sure that all other superparticles are
sufficiently heavy so that they do not play a role in the calculation
in the relic density. The precise values of their masses are then
irrelevant to us. We therefore fix the soft mass parameters of the
gauginos and sfermions to certain values well above
$M_{\tilde \nu_{R,1}}$; recall from eq.(\ref{neut1}) that this implies
an {\em upper} bound on the mass $M_4$ of the $U(1)'$ gaugino. As
noted in Sec.~\ref{section2} we set ${\bf Y_\nu} = 0$, since the small
values of the neutrino masses force them to be negligible for the
calculation of the relic density. We also set most of the scalar
trilinear couplings to zero, except the top trilinear coupling $T_t$
which we use together with $\tan\beta$ and $M_3$ to keep the SM Higgs
mass in the range $125 \pm 3$ GeV, where the uncertainty is dominated
by the theory error \cite{Carena:2013ytb}. Since we are interested in
superparticle masses in excess of $10$ TeV, the correct value of
$M_{h_1}$ can be obtained with a relatively small value of
$\tan\beta$, which we also fix.

As already noted in the previous Section, all relevant interactions
of $\tilde \nu_{R,1}$ scale (either linearly or quadratically) with 
the $U(1)'$ gauge coupling $g'$. Since our set--up is inspired by
gauge unification, we set this coupling equal to the $U(1)_Y$ coupling
in GUT normalization, i.e.
\begin{equation} \label{gprime}
g' = \sqrt{ \frac{5}{3}} g_1\,.
\end{equation}
Note also that the charges in Table~1 are normalized such that
$\sum \left( Q'_\psi \right)^2 = \sum \left( Q'_\chi \right)^2 =
\frac{3}{5} \sum Y^2$, where the sum runs over a complete 
$\mathbf{27}-$dimensional representation of $E_6$ \cite{London:1986dk}.
We will later comment on how the upper bound on $M_{\tilde \nu_{R,1}}$
changes when $g'$ is varied.

Recalling that we work in a basis where the matrix
${\mathbf m}^2_{\tilde N^C}$ is diagonal, with $m^2_{\tilde N^C,11}$
being its smallest element, the remaining relevant free parameters are
thus:
\begin{equation} \label{eq 4.1}
m^2_{\tilde{N}^C, 11}, \quad v_s, \quad  \lambda, \quad T_\lambda \quad \text{and} 
\quad  \theta_{E_6}\,.
\end{equation}
All these parameters are related to the extended sector that the UMSSM
has in addition to the MSSM. Since the mixing angle $\theta_{E_6}$
defines the $U(1)'$ gauge group, we want to determine the upper bound
on the mass of the lightest RH sneutrino as a function of
$\theta_{E_6}$. We will see below that this will also allow to derive
the absolute upper bound, valid for all versions of the UMSSM.

From the discussion of the previous Section we know that the first two
of the parameters listed in (\ref{eq 4.1}) are strongly correlated by
the requirement that $M_{\tilde \nu_{R,1}}$ is close to
$M_{Z'}/2$. More precisely, the minimal relic density is found if the
RH sneutrino mass is very roughly one $h_3$ decay width below the
nominal pole position, the exact distance depending on the couplings
$C_1$ and $C_2$; this shift from the pole position is due to the
finite kinetic energy of the sneutrinos at temperatures around the
decoupling temperature \cite{Griest:1990kh}.

The parameters $\lambda$ and $T_\lambda$ have to satisfy some bounds. First,
requiring the mass of the $SU(2)_{L}$ higgsinos to be at least $20\%$ larger
than $M_{Z'}/2$ leads to the lower bound
\begin{equation} \label{eq 4.2}
\lambda > 0.85  g' |Q'_S| \,,
\end{equation}
where we have used eqs.(\ref{eq 1.4}) and (\ref{eq 2.10}). Moreover,
$T_\lambda$ has to satisfy the upper bound (\ref{eq 3.2}), so that pairs of
the heavy $SU(2)_{L}$ doublet Higgs bosons can be produced in $\tilde \nu_{R,1}$
annihilation with $M_{\tilde \nu_{R,1}} \simeq M_{Z'}/2$. Having fixed $\tan\beta$ and $T_\lambda$, the effective final state
coupling $C_2$ defined in eq.(\ref{eq 3.8}) depends only on $\lambda$,
which is constrained by eq.(\ref{eq 4.2}); fortunately this still leaves
us enough freedom to vary $C_2$ over a sufficient range.

The bound on the lightest RH sneutrino mass for a given value of
$\theta_{E_6}$ can then be obtained as follows. We start by choosing
some value of $M_{h_3} \simeq M_{Z'}$ in the tens of TeV range. Note
that this fixes the coupling $C_1$, since we have already fixed $g'$
and $\theta_{E_6}$ and hence the charge $Q'_{N^C}$. We then minimize
the relic density for that value of $M_{h_3}$ by varying the
soft--breaking contribution to the sneutrino mass and $\lambda$; as
noted in Sec.~3, the minimum is reached when the physical RH sneutrino
mass is just slightly below $M_{Z'}/2$, and $C_2$ is close to the initial state
coupling $C_1$ of eq.(\ref{eq 3.3}). If the resulting relic density
$(\Omega h^2)_1$ is very close to the measured value of
eq.(\ref{relden}), we have found the upper bound on $M_{Z'}$ and hence
on $M_{\tilde \nu_{R,1}}$. Otherwise, we change the value of
$M_{h_3}$ by the factor $\sqrt{0.12 / (\Omega h^2)_1}$, and repeat the
procedure. Since the minimal relic density to good approximation
scales like $M_{h_3}^2$, see eq.(\ref{eq 3.1}), this algorithm
converges rather quickly.

\subsection{The $U(1)_{\psi}$ Model}  
\label{subsection4.2}

We illustrate our procedure first for $U(1)' = U(1)_\psi$, where the
$U(1)'$ charge of the RH sneutrinos is relatively small (in fact, the
same as for all SM (s)fermions). We choose the SUSY breaking scale to
be 18 TeV and we fix $\tan{\beta}=1.0$, $M_3=18$ TeV, and
${\bf m^2_{\tilde Q}} = {\bf m^2_{\tilde U^C}} = {\bf m^2_{\tilde
    D^C}} = 2 \times 10^8$ GeV$^2\cdot {\bf 1}$,
${\bf m^2_{\tilde L}}={\bf m^2_{\tilde E^C}} = 2.25 \times 10^{8}$
GeV$^2 \cdot {\bf 1}$,
$\left(m^2_{\tilde N^C}\right)_{22} = 2.2 \times 10^8$ GeV$^2$,
$\left(m^2_{\tilde N^C}\right)_{33} = 2.3 \times 10^{8}$ GeV$^2$. To
keep $M_{h_{1}}$ close to 125 GeV, the top trilinear coupling took
values in the following range ${T_{u,33}}= [-55, -33]$ TeV; recall
that the physical squared sfermion masses also receive $D-$term
contributions, which amount to $M^2_{Z'}/8$ in this model.

In this model the two Higgs doublets have the same $U(1)'$ charge, and
the product $Q'_{H_u} Q'_S$ is negative. As a result, the $\lambda^2$
and the $g'^2$ terms in the diagonal couplings given in eqs.(\ref{eq
  3.4}) and (\ref{eq 3.5}) tend to cancel, while the contribution
$\propto g'^2$ to the off--diagonal couplings given in eq.(\ref{eq
  3.6}) vanishes.  The contributions involving these off--diagonal
couplings are therefore subdominant. The largest contribution usually
comes from final states involving two heavy $SU(2)_{L}$ doublet Higgs
bosons, but the contributions from two light states (including the
longitudinal modes of the gauge bosons) are not much
smaller. Moreover, due to this cancellation we need relatively large
values of $\lambda$; the numerical results shown below have been
obtained by varying it in the range from $0.32$ to $0.46$.

\begin{figure}[h!]
\begin{subfigure}[b]{0.5\linewidth}
\centering\includegraphics[width=8.2cm, height=6cm]{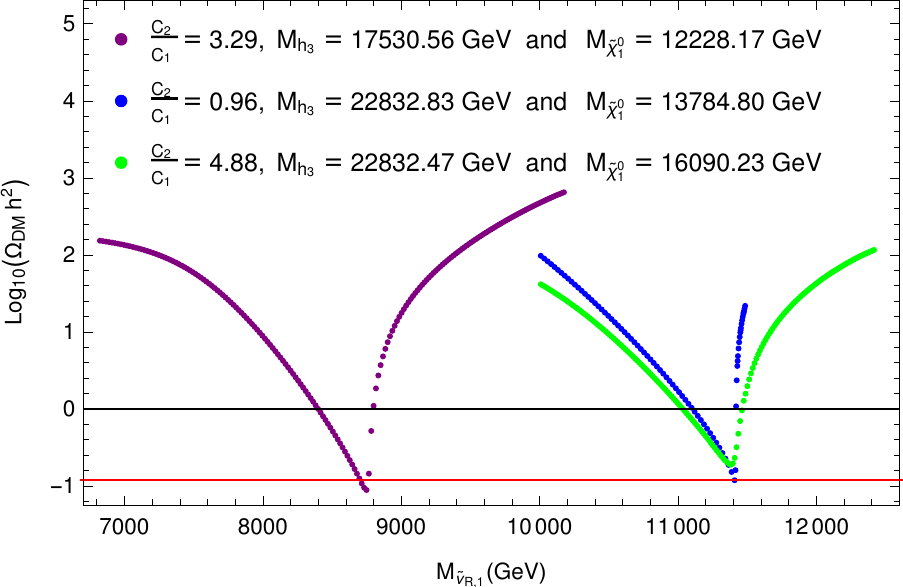}
\caption{}
\end{subfigure}\hfill
\begin{subfigure}[b]{0.58\linewidth}
\centering\includegraphics[width=7.0cm, height=6cm]{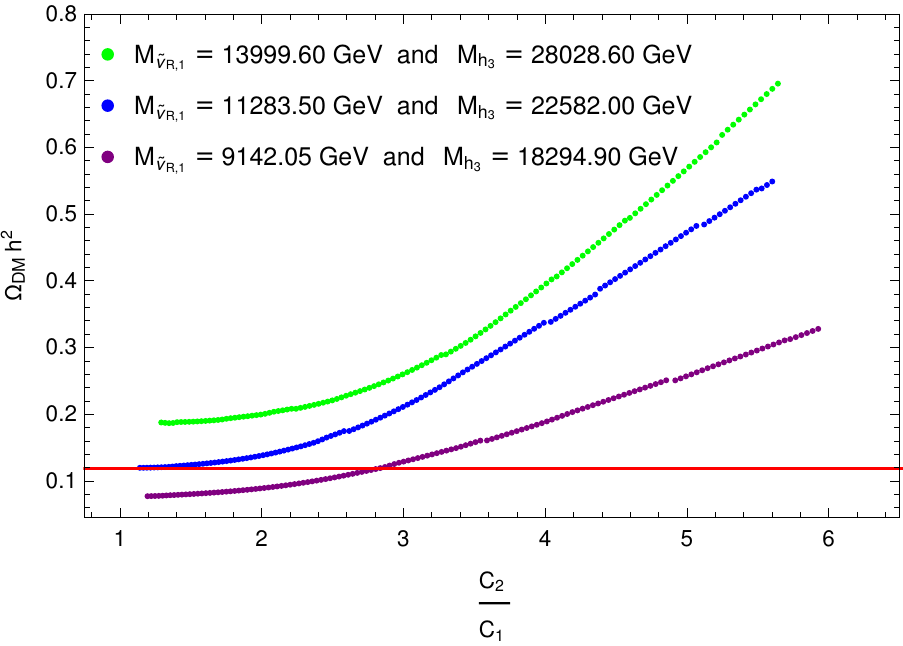}
\caption{}
\end{subfigure}   
\caption{Relic density as a function of
  $M_{\tilde{\nu}_{R,1}}$ (left) and its dependence on the ratio of
  couplings $\frac{C_{2}}{C_{1}}$ (right) for different singlet Higgs
  masses. The red lines correspond to the limits on the dark matter
  abundance obtained by the Planck Collaboration,
  $\Omega_{\text{DM}} h^2 = 0.1188\pm 0.0010$.
\label{Fig3}}
\end{figure}

Figure \ref{Fig3}a depicts the relic abundance of the RH sneutrino as
a function of $M_{\tilde{\nu}_{R,1}}$ for different values of the mass
of the singlet Higgs boson. All the curves show a pronounced minimum
when $M_{\tilde{\nu}_{R,1}}$ is very close to but below
$M_{h_{3}}/2$. The blue and the green curves are for $v_s = 59$ TeV
and thus have the same coupling $C_1$ and (approximately) the same
mass of the singlet Higgs, but the blue curve has a smaller value of
$C_2$. This reduces the width of $h_3$ as well as the annihilation
cross section away from the resonance, and therefore leads to a
narrower minimum.

In figure \ref{Fig3}b we show the dependence of the relic density on
the ratio of couplings $C_2 / C_1$ for fixed mediator masses close to
the resonance. This confirms our expectations from the previous
Section: if $C_2$ is significantly larger than $C_1$, the relic
density increases with $C_2$ because the increase of the mediator
decay width over--compensates the increased coupling strength in the
total annihilation cross section. If $C_{2}\ll C_{1}$ the width of the
mediator is dominated by mediator decays into
$\tilde \nu_{R,1} \tilde \nu^*_{R,1}$; hence increasing $C_2$ reduces
the relic density because it increases the normalization of the
annihilation cross section. Note that the relic density curve is
fairly flat over some range of $C_2 / C_1$. Moreover, the optimal
choice of $C_2/C_1$ also depends somewhat on how far
$M_{\tilde \nu_{R,1}}$ is below $M_{h_3}/2$. Altogether, for given
$M_{h_3}$ there is an extended $1-$dimensional domain in the
$(M_{\tilde \nu_{R,1}}, C_2/C_1)$ plane over which the relic density is
quite close to its absolute minimum. This simplifies our task of
minimization. Note also that we calculate the annihilation cross
section only at tree--level; a change of the predicted relic density
that is smaller than a couple of percent is therefore not really
physically significant.

The parameters of the blue curve in Fig.~\ref{Fig3}b in fact are
very close to those that maximize $M_{\tilde \nu_{R,1}}$ within the
$U(1)_\psi$ model, under the assumption that $\tilde \nu_{R,1}$ was in
thermal equilibrium in standard cosmology.
$M^{\rm max}_{\tilde{\nu}_{R,1}} \simeq 11.5$ TeV corresponds to an
upper bound on $M_{h_3}$ and $M_{Z'}$ of about $23.0$ TeV. This is
clearly beyond the reach of the LHC, and might even stretch the
capabilities of proposed $100$ TeV $pp$ colliders.

Recall that all left--handed SM (anti)fermions have the same
$U(1)_\psi$ charge. As a result, in the absence of $Z - Z'$ mixing the
$Z' f \bar f$ couplings are purely axial vector couplings, for all SM
fermions $f$. $Z'$ exchange can therefore only contribute to
spin--dependent WIMP--nucleon scattering in this model. Since our WIMP
candidate doesn't have any spin, $Z'$ exchange does not contribute at
all. Once $Z-Z'$ mixing is included, $Z$ exchange contributes a term
of order
$M_{\tilde \nu_{R,1}} M_N \sin \alpha_{ZZ'} / M_Z^2 \propto M_{\tilde
  \nu_{R,1}} M_N / M_{Z'}^2$
to the matrix element for $\tilde \nu_{R,1} N$ scattering, while the
mixing--induced $Z'$ exchange contribution is suppressed by another
factor $M_Z^2 / M_{Z'}^2$; here $M_N$ is the mass of the
nucleon. There is also a small contribution from the light SM--like
Higgs boson $h_1$, which is very roughly of order $M_N^2 /
M_{h_1}^2$. As a result the scattering cross section on nucleons is very small,
below $10^{-13}$ pb for the scenario that maximizes
$M_{\tilde \nu_{R,1}}$. For the given large WIMP mass, this is not
only several orders of magnitude below the current bound, but also
well below the background from coherent neutrino scattering
(``neutrino floor'').

\subsection{The $U(1)_\eta$ Model} 
\label{subsection4.3}

We now consider a value of $\theta_{E_6}$ with a larger $U(1)'$ charge
of the right--handed neutrino superfields. This increases the coupling
$C_1$ for given $M_{Z'}$, and thus the $\tilde \nu_{R,1}$ annihilation
cross section for given masses, which in turn will lead to a weaker
upper limit on $M_{\tilde \nu_{R,1}}$ from the requirement that the
$\tilde \nu_{R,1}$ relic density not be too large.

In our analysis we therefore choose the SUSY breaking scale to be 50
TeV and we fix $\tan\beta = 2.2$, and
${\bf m^2_{\tilde Q}} = 1.28 \times 10^9$
GeV$^{2} \cdot{\bf 1}, {\bf m^2_{\tilde U^C}} = 1.45 \times 10^9$
GeV$^{2} \cdot{\bf 1}$, ${\bf m^2_{\tilde D^C}} = 3.0 \times 10^9$
GeV$^{2} \cdot {\bf 1}$, ${\bf m^2_{\tilde L}} = 3.0 \times 10^9$
GeV$^{2} \cdot{\bf 1}$, ${\bf m^2_{\tilde E^C}} = 1.28 \times 10^9$
GeV$^{2} \cdot {\bf 1}$,
$\left(m^2_{\tilde N^C}\right)_{22} = -4.0 \times 10^8$ GeV$^2$,
$\left(m^2_{\tilde N^C}\right)_{33} = -3.9 \times 10^8$ GeV$^2$. To
keep $M_{h_{1}}$ close to 125 GeV, the top trilinear coupling took
values in the range ${T_{u,33}}= [-130, -114]$ TeV.  In this case the
$U(1)'$ $D-$term contributions are positive for
$\tilde Q, \, \tilde u^C$ and $\tilde \nu_R$, but are negative for
$\tilde L$ and $\tilde d^C$.

In this model the two Higgs doublets have different $U(1)'$ charges;
hence there is a sizable gauge contribution to the off--diagonal
couplings of eq.(\ref{eq 3.6}). The Higgs doublet charges again have
the opposite sign as the charge of $S$, leading to cancellations
between the $\lambda^2$ and $g'^2$ terms in the diagonal couplings
(\ref{eq 3.4}) and (\ref{eq 3.5}). This cancellation is particularly
strong for the coupling to two light states, so that for the
interesting range of $\lambda$ the most important final states involve
two heavy $SU(2)_{L}$ doublets, although final states with one light and
one heavy boson are also significant. Partly because of this, and partly
because the coefficients of the $g'^2$ terms are smaller than in the
$U(1)_\psi$ model, smaller values of the coupling $\lambda$ are
required; the numerical results below have been obtained with
$\lambda \in [0.260, 0.352]$.
  
\begin{figure}[h!]
\begin{subfigure}[b]{0.5\linewidth}
\centering\includegraphics[width=7.8cm, height=6cm]{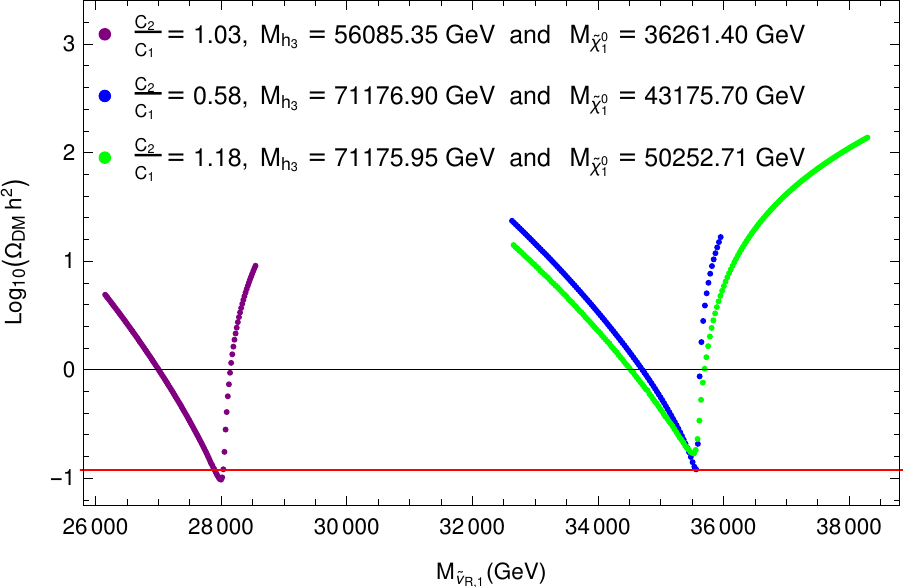}
\caption{}
\end{subfigure}\hfill
\begin{subfigure}[b]{0.55\linewidth}
\centering\includegraphics[width=7.0cm, height=6cm]{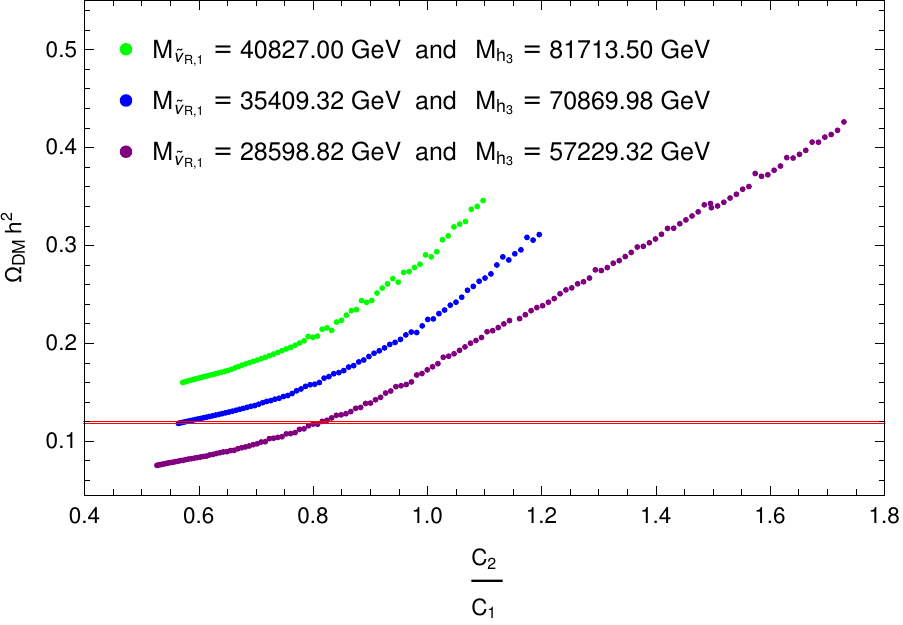}
\caption{}
\end{subfigure}   
\caption{As in Fig.~\ref{Fig3}, but for the $U(1)_\eta$ model. \label{Fig4}}
\end{figure}

In Fig.~\ref{Fig4} we again show the dependence of the relic density on
the mass of the lightest RH sneutrino (left) and on the ratio of couplings
$C_2/C_1$ (right). The qualitative behavior is similar to that in
the $U(1)_\psi$ model depicted in Fig.~\ref{Fig3}, but clearly much larger
values of $M_{\tilde \nu_{R,1}}$ are now possible, the absolute upper bound
being near $35$ TeV (see the blue curves). The corresponding $Z'$ mass
of about $70$ TeV is definitely beyond the reach of a $pp$ collider 
operating at $\sqrt{s} = 100$ TeV

Since $Q'_Q = Q'_{U^C} \neq Q'_{D^C}$ in this model, there is no
vector coupling of the $Z'$ to up quarks, but such a coupling does
exist for down quarks. Hence now the $Z'$ exchange contribution to the
matrix element for elastic scattering of $\tilde \nu_{R,1}$ on
nucleons is comparable to that of $Z$ exchange once $Z - Z'$ mixing
has been included, and the $h_1$ exchange contribution has roughly the
same size as in the $U(1)_\psi$ model. The total $\tilde \nu_{R,1} N$
scattering cross sections are again below $10^{-13}$ pb, for
parameters near the upper bound on $M_{\tilde \nu_{R,1}}$. Since our
WIMP candidate is now even heavier than in the $U(1)_\psi$ model, this
is even more below the current constraints as well as below the
neutrino floor.

\subsection{The General UMSSM}  
\label{subsection4.4}

In this subsection we investigate in more detail how the upper bound
on $M_{\tilde \nu_{R,1}}$ depends on $\theta_{E_6}$. To this extent we
have applied the procedure outlined in subsec.~\ref{subsection4.1},
and applied to two specific $U(1)'$ models in subsecs.~\ref{subsection4.2}
and \ref{subsection4.3}, to several additional $U(1)'$ models, each
with a different value of $\theta_{E_6}$.

\begin{figure}[h!]
\centering
\includegraphics[width=0.8\textwidth]{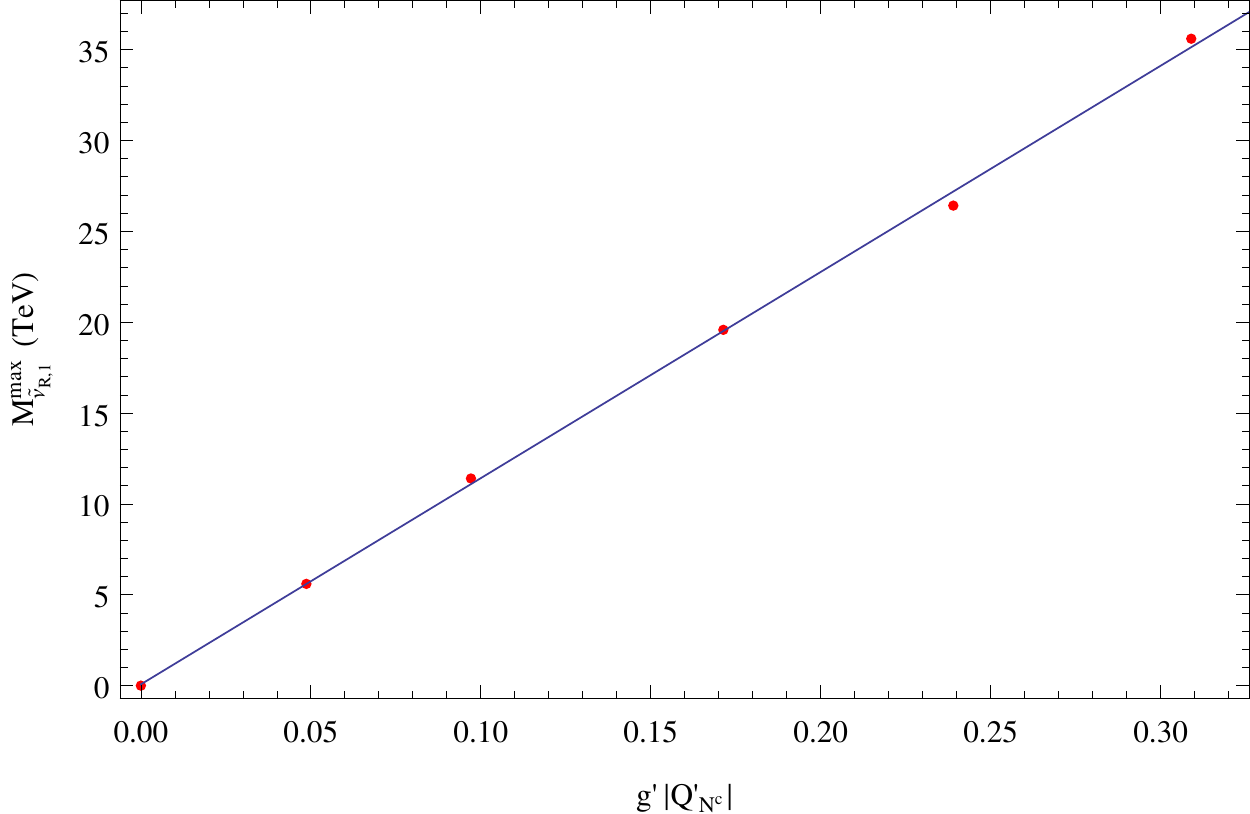}
\caption{The upper limit on $M_{\tilde \nu_{R,1}}$ derived from the
relic density as a function of $\vert Q'_{N^C} \vert$. The straight line shows a linear fit to the six numerical results.
\label{Fig5} }
\end{figure}

The results are shown in Fig.~\ref{Fig5}, where we plot the upper bound
on the mass of the lightest RH sneutrino as a function of the absolute value
of the product $g' Q'_{N^C}$. In order of increasing $\vert Q'_{N^C} \vert$, the 
six red points correspond to the following choices of $\theta_{E_{6}}$:
\[
\\ \biggl\{ \arctan\sqrt{15}, \frac{ \left( \arctan\sqrt{15} + \frac{\pi}{2}
\right) } {2}, \frac{\pi}{2}, \frac{ \left( \arctan\sqrt{\frac{3}{5}} + 
\arctan \left[ \frac{7} {\sqrt{15}} \right] \right) } {2} ,
\arctan\sqrt{\frac{3}{5}}, -\arctan\sqrt{\frac{5}{3}} \biggr\}\,.
\]
Note that the first point has a vanishing $U(1)'$ charge for the $N^C$
superfields, i.e. the resonance enhancement of the annihilation cross
section does not work in this case.  We checked that the cross section
for elastic $\tilde \nu_{R,1} N$ scattering is well below the
experimental bound for all other points.

Evidently the upper bound on $M_{\tilde \nu_{R,1}}$ scales essentially
linearly with $Q'_{N^C}$; recall that $g'$ has been fixed to
$\sqrt{5/3} g_1$ here. This linear dependence can be understood as
follows. The $h_3 \tilde \nu_{R,1} \tilde \nu_{R,1}^*$ coupling can be
written as $g' Q'_{N^C} M_{Z'} \simeq 2 g' Q'_{N^C} M_{\tilde \nu_{R,1}}$.
Moreover, we saw above that the maximal sneutrino mass is allowed if
the effective final--state coupling $C_2$ is similar to $C_1$; it is
therefore also proportional to $Q'_{N^C}$. Therefore at the point
where the bound is saturated, the $h_3$ decay width scales like
$|C_1|^2 M_{h_3} \propto g'^2 (Q'_{N^C})^2 M_{\tilde \nu_{R,1}}$,
where we have again used that near the resonance all relevant masses
are proportional to $M_{\tilde \nu_{R,1}}$. Note finally that for a
narrow resonance -- such as $h_3$, for the relevant parameter choices
-- the thermal average over the annihilation cross section scales like
$1/(M_{h_3} \Gamma_{h_3})$ \cite{Griest:1990kh}. Altogether we thus have
\begin{equation} \label{scaling}
\langle \sigma v \rangle \propto \frac{ |C_1 C_2|^2 } { M_{h_3}
\Gamma_{h_3} M^4_{\tilde \nu_{R,1}} }
\propto  \frac {g'^2 (Q'_{N^C})^2} {M^2_{\tilde \nu_{R,1}}} \,.
\end{equation}
The linear relation between the upper bound on $M_{\tilde \nu_{R,1}}$
and $Q'_{N^C}$ then follows from the fact that the thermally averaged
annihilation cross section essentially fixes the relic density.

Note that here $Q'_{N^C}$ always comes with a factor $g'$; indeed, for a
$U(1)$ gauge interaction only the product of gauge coupling and charge
is well defined. The linear dependence of the bound on $M_{\tilde \nu_{R,1}}$
on $Q'_{N^C}$ for fixed $g'$ depicted in Fig.~\ref{Fig5} can therefore also be
interpreted as linear dependence of the bound on the product $g' Q'_{N^C}$.
A fit to the points in Fig.~\ref{Fig5} gives:
\begin{equation} \label{fit}
M^{\rm max}_{\tilde \nu_{R,1}} = ( 0.071 + 113.477 g' |Q'_{N^C}| ) \
{\rm TeV}\,.
\end{equation}
This is the central result of our paper.

The highest absolute value of $|Q'_{N^{c}}|$ in the UMSSM is about
$0.82$, which is saturated for
$\theta_{E_6} = -\arctan\left[\frac{1}{\sqrt{15}}\right]$. Using the
linear fit of eq.(\ref{fit}) and $g' = \sqrt{5/3} g_1 = 0.47$ leads to
an absolute upper bound on $M_{\tilde \nu_{R,1}}$ in unifiable
versions of the UMSSM of about $43.8$ TeV. This corresponds to an
absolute upper bound on the $Z'$ mass of about $87.6$ TeV.

Finally, we recall from eq.(\ref{th_range}) that for $\theta_{E_6}$
between $-\arctan\sqrt{15}$ and $0$ one needs a negative squared soft
breaking mass in order to have $M_{\tilde \nu_{R,1}} \simeq M_{Z'}/2$.
Since the $\hat N^C$ superfields appear in the superpotential (\ref{eq
  1.3}) only multiplied with the tiny couplings ${\bf Y_\nu}$, this
superpotential will not allow to generate negative squared soft
breaking masses for sneutrinos via renormalization group running
starting from positive values at some high scale. If we insist on
positive squared soft breaking mass for all $\tilde \nu_R$ fields the
upper bound on $|Q'_{N^C}|$ is reduced to $\sqrt{5/8} \simeq 0.79$, in
which case the bound on $M_{\tilde \nu_{R,1}}$ is reduced to about
$42$ TeV. We note, however, that the $\hat N^C$ superfields can have
sizable couplings to some of the exotic color triplets that reside in
the ${\mathbf 27}-$dimensional representation
\cite{Hewett:1988xc}. Recalling that at least some of these exotic
fermions are usually required for anomaly cancellation it should not
be too difficult to construct a UV complete model that allows negative
squared soft breaking terms for (some) $\tilde \nu_R$ at the SUSY mass
scale.

\subsection{Prospects for Detection}
\label{section4.5}

Clearly spectra near the upper bound presented in the previous subsection
are not accessible to searches at the LHC, nor even to a proposed 100 TeV
$pp$ collider.

As already noted for the $U(1)_\eta$ and $U(1)_\psi$ models the
$\tilde \nu_{R,1}$ nucleon scattering cross section is very small. The
very large $Z'$ mass suppresses the $Z'$ exchange contribution; as we saw
in Sec.~\ref{section2.3} it also suppresses $Z-Z'$ mixing, so that the
$Z$ exchange contribution also scales like $M_{Z'}^{-2}$. The
contribution due to the exchange of the singlet--like Higgs boson
($h_3$ in our analysis) is suppressed by the very large value of
$M_{h_3}$ as well as the tiny $h_3 q \bar q$ couplings, which solely
result from mixing between singlet and doublet Higgs bosons. Finally,
the contribution from the exchange of the doublet Higgs bosons, in
particular of the 125 GeV state $h_1$, is suppressed by the small size
of the $h_1 \tilde \nu_{R,1} \tilde \nu_{R,1}^*$ coupling, which is of
order $g' v \ll M_{\tilde \nu_{R,1}}$, as well as the rather small
$h_1 q \bar q$ couplings, which are much smaller than gauge
couplings. As a result, the $\tilde \nu_{R,1}$ nucleon scattering
cross section, and hence the signal rate in direct WIMP detection
experiments, is well below the neutrino--induced background; recall
that this ``neutrino floor'' increases $\propto M_{\tilde \nu_{R,1}}$
since the WIMP flux, and hence the event rate for a given cross
section, scales $\propto 1/M_{\tilde \nu_{R,1}}$.

The best chance to test these scenarios therefore comes from indirect
detection. Naively one expects the cross section for annihilation from
an $S-$wave initial state to be essentially independent of
temperature, in which case the correct thermal relic density implies
$\langle \sigma v \rangle \simeq 2.4 \cdot 10^{-26} {\rm \text{ cm}}^3/{\rm \text{s}}
\simeq 0.8 {\rm \text{ pb}}\cdot \text{c}$ \cite{Steigman:2012nb, Drees:2015exa}.
However, as pointed out in \cite{Ibe:2008ye, Guo:2009aj} this can
change significantly in the resonance region; here the thermally
averaged annihilation cross section can be significantly higher in
today's universe than at the time of WIMP decoupling.

\begin{figure}[h!]
\centering
\includegraphics[width=0.8\textwidth]{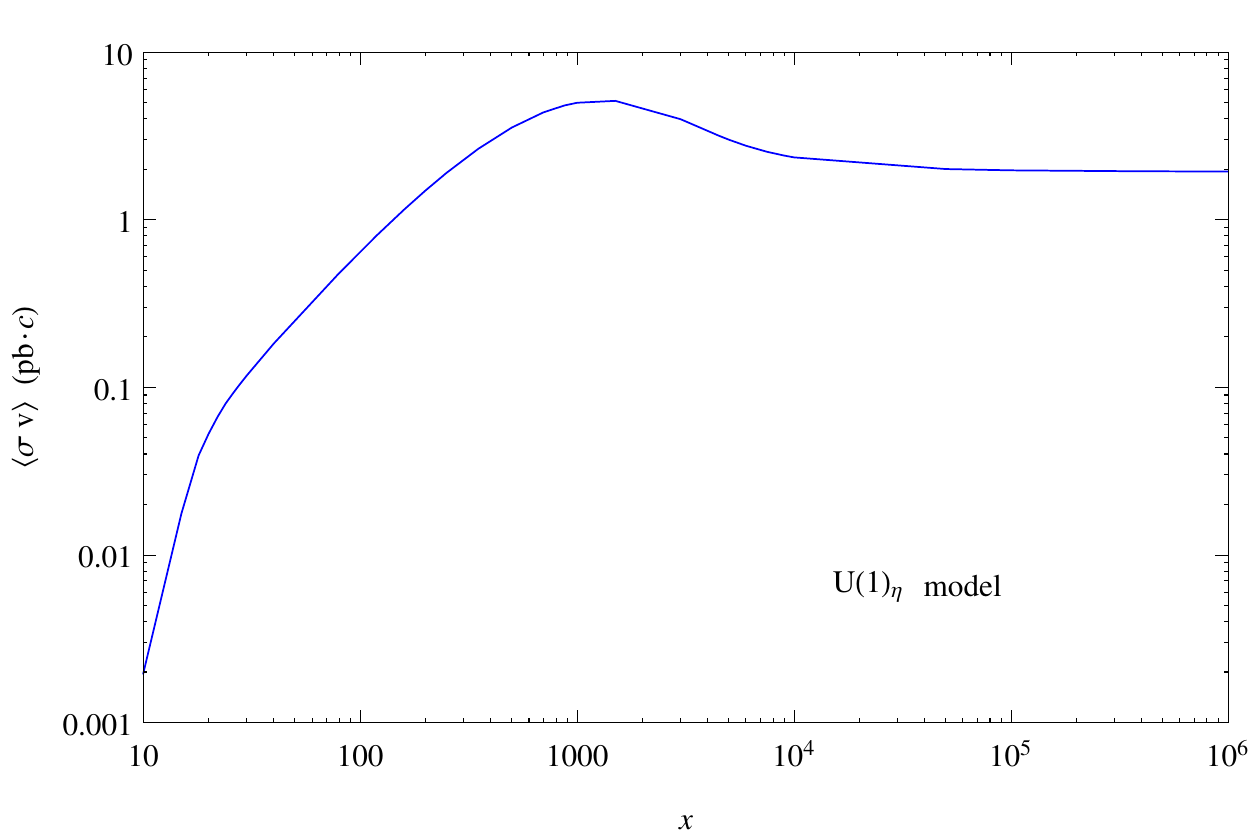}
\caption{Thermally averaged cross section as a function of the scaled
  inverse temperature $x \equiv M_{\tilde \nu_{R,1}}/T$ for the parameters
  of the $U(1)_\eta$ model that saturate the upper bound on the sneutrino
  mass. Nominal decoupling occurs at $x=x_F=27.2$, whereas in today's
  galaxies $x \sim 10^6$.
\label{Fig6} }
\end{figure}

This is illustrated in Fig.~\ref{Fig6} for the parameter choice that
saturates the upper bound on $M_{\tilde \nu_{R,1}}$ in the $U(1)_\eta$
model.  Here we show the thermally averaged
$\tilde \nu_{R,1} \tilde \nu_{R,1}^*$ annihilation cross section times
relative velocity as function of the scaled inverse temperature
$x = M_{\tilde \nu_{R,1}}/T$.\footnote{The total $\tilde \nu_{R,1}$
  annihilation rate also receives a contribution from
  $\tilde \nu_{R,1} \tilde \nu_{R,1} \rightarrow \nu \nu$ annihilation
  via neutralino exchange in the $t-$ and $u-$channels. However, since
  this contribution is not resonantly enhanced, it can safely be
  neglected.} We see that for a quite extended range of temperatures
around the decoupling temperature, $\langle \sigma v \rangle$ grows
almost linearly with $x$. This is because $M_{\tilde \nu_{R,1}}$ is
only slightly below the nominally resonant value $M_{h_3}/2$; by
reducing the temperature the fraction of the velocity distribution
that falls within approximately one $h_3$ decay width of the pole
therefore at first increases.

Today's relic density is essentially inversely proportional to the
``annihilation integral'', defined as \cite{Griest:1990kh}
\begin{equation} \label{J}
  J(x_F) = \int_{x_F}^\infty \frac { \langle \sigma v \rangle } {x^2} dx\,.
\end{equation}
An annihilation cross section that grows significantly for $x > x_F$
therefore has to be compensated by a smaller value of $\langle \sigma
v \rangle(x_F)$ in order to keep the relic density constant. As a result,
in our scenarios the annihilation cross section at decoupling is actually
significantly {\em smaller} than for typical $S-$wave annihilation.

Because for parameters that saturate the upper bound on $M_{\tilde{\nu}_{R,1}}$ the right-handed sneutrino mass is
somewhat below $M_{h_3}/2$, for very large $x$, i.e. very small
temperature, the thermally averaged annihilation cross section starts to
decrease again. However, for the parameters of Fig.~\ref{Fig6} it
asymptotes to a value that is still about three times larger than the
``canonical'' thermal WIMP annihilating from an $S-$wave initial state. As
shown in refs. \cite{Ibe:2008ye, Guo:2009aj} this enhancement factor
strongly depends on $2 M_{\tilde \nu_{R,1}} - M_{h_3}$; it can be even larger
for slightly smaller sneutrino masses that are even closer to $M_{h_3}/2$.

The WIMP annihilation rate in today's universe scales like the square
of the WIMP number density. This means that the flux of annihilation
products scales like $1/M^2_{\tilde \nu_{R,1}}$; for parameters
(nearly) saturating our upper bound on the sneutrino mass it is thus
too small to be detectable by space--based observatories like FermiLAT
\cite{Ackermann:2015zua}, simply because of their small size.  Recall
also that our sneutrinos annihilate into (longitudinal) gauge or Higgs
bosons, and thus mostly into multi--hadron final states. This leads to
a continuous photon spectrum which, for parameters near the upper
bound on the sneutrino mass, extends well into the TeV region. Photons
of this energy can be detected by Cherenkov telescopes on the ground,
via their air showers. Note also that the astrophysical cosmic ray
background drops even faster than $E^{-2}$ with increasing energy $E$
of the cosmic rays; the signal to background ratio therefore actually
improves with increasing WIMP mass. Indeed, simulations show that at
least for a favorable distribution of dark matter particles near the
center of our galaxy, the continuum photon flux of multi--TeV WIMPs
annihilating with the canonical thermal cross section should be
detectable by the Cherenkov Telescope Array \cite{Carr:2015hta}.

\section{Summary and Conclusions}  
\label{section5}

In this paper we analyzed the UMSSM, i.e. extensions of the minimal
supersymmetrized Standard Model that contain an additional $U(1)'$
gauge group as well as additional right--handed (RH) neutrino
superfields which are singlets under the SM gauge group but carry
$U(1)'$ charge. We assume that $U(1)'$ is a subgroup of $E_6$, which
has been suggested as an (effective) GUT group, e.g. in the context of
early superstring phenomenology. In this case the lightest RH
sneutrino $\tilde \nu_{R,1}$ can be a good dark matter candidate.

We found that even within minimal cosmology, and fixing the $U(1)'$
gauge strength to be equal to that of the hypercharge interaction of
the (MS)SM (in GUT normalization), $\tilde \nu_{R,1}$ masses of tens
of TeV are possible.  For given $U(1)'$ charges the bound on
$M_{\tilde \nu_{R,1}}$ is saturated if $\tilde \nu_{R,1}$ can
annihilate resonantly through the exchange of both the new $Z'$ gauge
boson and of the new Higgs boson $h_3$ associated with the spontaneous
breaking of $U(1)'$; note that $M_{Z'} \simeq M_{h_3}$ {\em
  automatically} in this model. Scalar $h_3$ exchange is more
important since $Z'$ exchange can only occur from a $P-$wave initial
state. The $h_3 \tilde \nu_{R,1} \tilde \nu_{R,1}^*$ coupling is
fixed by the $U(1)'$ charge $Q'_{N^C}$ of the right--handed neutrinos,
but the $h_3$ couplings to the relevant final states can be tuned
independently, allowing a further maximization of the annihilation
cross section. In our analysis we used $SU(2)_{L}$ doublet Higgs
bosons as well as longitudinal $W$ and $Z$ bosons as final
states. While the light $SU(2)$ doublet Higgs states, including the
longitudinal $W$ and $Z$ modes, are always accessible, we could have
replaced the heavy Higgs doublet in the final state by some exotic
fermions which in most cases are required to cancel anomalies. The
only requirement is that the effective final state coupling of $h_3$
should be tunable to values close to its coupling to
$\tilde \nu_{R,1}$. Since the $Z'$ exchange contribution is basically
fixed by $\theta_{E_6}$, and non--resonant contributions are negligible
for $M_{\tilde\nu_{R,1}} \sim M_{Z'}/2$, most of the many free
parameters of this model, which describe the sfermion and gaugino
sectors, are essentially irrelevant to us. The only requirement is
that these superparticles are sufficiently heavy to avoid
co--annihilation, which would increase the relic density in our case.

We found that the final upper bound on $M_{\tilde \nu_{R,1}}$ is
essentially proportional to the product $g' |Q'_{N^C}|$, where $g'$ is
the $U(1)'$ gauge coupling. Within the context of theories unifiable
into $E_6$ this leads to an absolute upper bound on
$M_{\tilde \nu_{R,1}}$ of about $43.8$ TeV. In other words, in this
fairly well motivated set--up we can find a thermal WIMP candidate
with mass less than a factor of three below the bound derived from
unitarity \cite{Griest:1989wd}. This is to be contrasted with an upper
bound on the mass of a neutralino WIMP in the MSSM of about 8 TeV for
unsuppressed co--annihilation with gluinos \cite{Ellis:2015vaa}. In a
rather more exotic model featuring a WIMP residing in the quintuplet
representation of $SU(2)$ a WIMP mass of up to $9.6$ TeV is allowed
\cite{Cirelli:2009uv}. 

Of course, this mechanism requires some amount of finetuning: the mass
of the WIMP needs to be just below half the mass of the $s-$channel
mediator. We find that typically the predicted WIMP relic density
increases by a factor of $2$ when the WIMP mass is reduced by between
$1$ and $3\%$ from its optimal value. In contrast, the recent proposal
to allow thermal WIMP masses near $100$ TeV via non--perturbative
co--annihilation requires finetuning to less than $1$ part in $10^5$
\cite{coan_new}.

We also note that our very heavy WIMP candidates have very small
scattering cross sections on nuclei, at least two orders of magnitude
below the neutrino floor. This shows that both collider searches and
direct WIMP searches are still quite far away from decisively probing
this reasonably well motivated WIMP candidate. On the other hand, we
argued that indirect signals for WIMP annihilation might be detectable
by future Cherenkov telescopes. Our analysis thus motivates extending
the search for a continuous spectrum of photons from WIMP annihilation
into the multi--TeV range.

While the result (\ref{fit}) has been derived within UMSSM models that
can emerge as the low--energy limit of $E_6$ Grand Unification, it
should hold much more generally. To that end $g' |Q'_{N^C}|$ should be
replaced by $g_{\chi \chi \phi}/m_\phi$, where $\chi$ is a complex
scalar WIMP annihilating through the near resonant exchange of the
real scalar $\phi$, $g_{\chi \chi \phi}$ being the (dimensionful)
$\chi \chi^* \phi$ coupling. In order to saturate our bound the
couplings of $\phi$ to the relevant final states should be tunable
such that the effective final state coupling, which we called $C_2$ in
Sec.~\ref{section3}, should be comparable to the initial--state
coupling $g_{\chi \chi \phi}$. In this case the algorithm we used to
find the upper limit on $M_{\tilde \nu_{R,1}}$, see
subsec.~\ref{subsection4.1}, can directly be applied to finding the
upper bound on $M_\chi$. We finally note that $M_\chi$ can be
increased by another factor of $\sqrt{2}$ if $\chi$ is a real scalar.

\acknowledgments

We thank Florian Staub and Manuel Krauss for useful explanations of
the usage of {\tt SARAH} {\tt SPheno}. This work was supported in part
by the Deutsche Forschungsgemeinschaft via the TR33 ``The Dark
Universe'', and in part by the Brazilian \emph{Coordination for the
  Improvement of Higher Education Personnel} (CAPES). Felipe A. Gomes
Ferreira would also like to thank the \emph{Bethe Center for
  Theoretical Physics (University of Bonn)} for the hospitality.

\end{document}